# When words collide: Bayesian meta-analyses of distractor and target properties in the picture-word interference paradigm

Audrey Bürki[1*], F.-Xavier Alario[2,3], & Shravan Vasishth[1]

[1]University of Potsdam, Karl-Liebknecht-Straße 24-25, 14476 Potsdam, Germany

[2] Aix Marseille Université, CNRS, LPC UMR 7290, Marseille, France

3 Department of Neurological Surgery, School of Medicine, University of Pittsburgh, USA

## Abstract

In the picture-word interference paradigm, participants name pictures while ignoring a written or spoken distractor word. Naming times to the pictures are slowed down by the presence of the distractor word. The present study investigates in detail the impact of distractor and target word properties on picture naming times, building on the seminal study by Miozzo and Caramazza (2003) "When more is less: A counterintuitive effect of distractor frequency in the picture-word interference paradigm. Journal of Experimental Psychology. General." We report the results of several Bayesian meta-analyses, based on 26 datasets. These analyses provide estimates of effect sizes and their precision for several variables and their interactions. They show the reliability of the distractor frequency effect on picture naming latencies (latencies decrease as the frequency of the distractor increases) and demonstrate for the first time the impact of distractor length, with longer naming latencies for trials with longer distractors. Moreover, distractor frequency interacts with target word frequency to predict picture naming latencies. The methodological and theoretical implications of these findings are discussed.

*Corresponding author (buerki@uni-potsdam.de)

## Introduction

A primary goal of psycholinguistics is to uncover the cognitive architecture underlying the production and comprehension of language. To do so, scientists investigate the representations and processes involved in language processing by designing, conducting, and interpreting tightly controlled experiments. Such experiments are necessary because of the many parameters that may affect verbal performance, and because introspection about the underlying mental processes has only a limited ability to reveal the representations and processes of interest.

In psycholinguistic research on language production, many studies have used the picture-word interference paradigm. The task consists in presenting stimuli that will trigger the production of words or sentences, in the context of other verbal stimuli, which are to be ignored when performing the task. Naming a picture in the context of a spoken or written distractor word creates interference, that is, it takes more time than naming a picture in the context of a non-meaningful stimulus. The picture-word interference paradigm was first used to study (incidental) reading abilities (e.g., Briggs & Underwood, 1982; Rayner & Posnansky, 1978; Underwood & Briggs, 1984, see also Alario et al., 2007) and has since then been applied to a variety of issues, such as the mechanisms of lexical access during word production (e.g., Abdel Rahman & Aristei, 2010; Damian & Bowers, 2003; Roelofs, 1992; Starreveld et al., 2013), the involvement of inhibition in language production (e.g., Shao et al., 2015), the scope of advanced planning (e.g., Meyer, 1996; Michel Lange & Laganaro, 2014), or the role of emotions in language processing (White et al., 2016). On the *Web of Science,* 541 articles are listed under the key word *picture-word interference* with a total number of citations (excluding self-citations) of 18271 during the two decades spanning between 1999 and 2018.

Such popularity of the picture-word interference paradigm is largely due to the assumption that it can provide precise information on the cognitive processes underlying word production and their time course (e.g., Levelt et al., 1999; Roelofs, 1992; Schriefers et al., 1990). The capability of the picture word-interference paradigm to tap into specific processes comes at a cost, however. The task requires the simultaneous processing of two types of stimuli, the picture to be named, and the incidental and presumed automatic processing of a written or auditory distractor word that must be ignored by instruction. Moreover, the instruction to name one of the stimuli and to ignore the other likely requires attentional or inhibition processes (e.g., Carr, 1999; Piai et al., 2011; Piai et al., 2012) that are not necessarily present in more ecological language production tasks. Numerous studies have attempted to determine how these processes combine and organize in time to generate the experimental patterns that we observe. The great enthusiasm generated by the picture-word interference paradigm has led to a myriad of empirical studies designed and performed to test increasingly refined theoretical questions. Over the years, new empirical results and the discrepancies across studies have been used to propose novel accounts or further refine existing ones at various levels of description, including computational accounts such as Weaver++ (Roelofs, 2003; Roelofs, 2018). To date however, no consensus has been reached (e.g., Miozzo & Caramazza, 2003; Starreveld et al., 2013). In the present study, we investigate the impact on picture naming processes of the properties of distractor words, most notably the unexpected distractor frequency effect.

In visual lexical decision or word naming tasks, more frequent words are processed faster (e.g., Ferrand et al., 2018). Miozzo and Caramazza (2003) set out to examine whether in a picture-word interference task, too, the frequency of the (written) distractor that had to be ignored would influence performance, where performance is quantified as the time needed to retrieve and produce the name of the picture. Asking this question was deemed an original and important test (see also Burt, 1994 and Burt, 2002) of the most canonical hypothesis describing lexical retrieval. Under the assumption that lexical access is a competitive process, where the time to name a picture depends on the activation level of the word's representation as well as on the activation levels of other lexical representations in the system (Levelt et al., 1999; Roelofs, 1992), these authors predicted that high frequency words would create more interference. Miozzo and Caramazza (2003) observed the reverse, that is, naming times were about 16 to 48 ms longer (depending on the experiment) for trials with distractors of lower frequency. This "reverse" distractor frequency effect has been replicated several times by several groups (e.g., Dhooge et al., 2013; Dhooge & Hartsuiker, 2010; Geng et al., 2014; Riès et al., 2015; Starreveld et al., 2013; de Zubicaray et al., 2012). The effect appears reliable, but its underlying mechanisms are still debated. To shed light on this issue, studies have examined the influence of other properties of the distractors on picture naming latencies, such as case alternation (Miozzo & Caramazza, 2003) as well as interactions between the frequency of the distractor and other variables, including target word frequency and semantic interference. Contrary to the main effect of distractor frequency, the results regarding these interactions appear to be inconsistent across studies.

In the present paper we build on Miozzo and Caramazza's (2003) seminal study and take advantage of data collected in the last three decades. The large quantities of picture-word interference datasets available can be (re-)used to perform meta-analyses of previously reported effects, or of effects pertaining to novel predictions. The outcome of a meta-analysis is an estimate of the size of an effect and of the uncertainty of the effect estimate. In a first meta-analysis, we quantify the size of the distractor frequency effect in experiments where this variable was *not* manipulated. We assess how much frequency, treated as a continuous measure, influences the naming times to the target words, when the impact of other variables that are often correlated with frequency is considered. We thereby obtain information about the size of the distractor frequency effect and of the precision of this estimate. In addition, we compute quantitative meta-analytic estimates of several main effects and interactions pertaining to other properties of the distractor words, to properties of the target words, or to the relationship between distractor and target words. The selection of effects is partly constrained by the available experiments' materials and, in particular, by the variables that were manipulated or can be computed in these datasets. In the remainder of this Introduction, we first discuss the variables examined in the present work. We articulate how they can contribute to a better understanding of the mechanisms underlying the distractor frequency effect and, more generally, to the mechanisms involved in picture-word interference tasks. We then briefly discuss the methodological implications of these analyses.

*Variables examined in the present study and their theoretical relevance*

From a theoretical perspective, knowledge about the variables that influence naming times in the picture-word interference paradigm may inform the mechanisms underlying participants' performance in this task. Whereas the distractor frequency effect itself appears robust, current accounts of this effect often rely on experimental effects and interactions - or the absence thereof- whose replicability has not yet been established. The empirical effects examined in the present study will not resolve all debates but will shed light on effects and interactions that have been deemed crucial to better understand the functional origin of the distractor frequency effect, or the conditions under which this effect arises.

We first examine the role of *distractor length*, a variable known to influence reading times but whose impact on picture naming latencies has seldom been considered. With this variable, we test the hypothesis that naming times in the picture-word interference paradigm partly depend on distractor processing duration. Miozzo and Caramazza (2003) discussed this hypothesis as a potential explanation for the distractor frequency effect, which they termed the *input* account. In this account, the distractor frequency effect arises because the distractor takes up some of the processing resources that would normally be used by the word production process. The longer the processing of the distractor, the longer the naming times. They reasoned that under such an account, any variable impacting the duration of written word processing should impact the picture naming times. Miozzo and Caramazza (2003) compared trials with distractors presented in alternating versus non-alternating case, or with distractors that had or had not been previously presented in a lexical decision task. None of these manipulations influenced naming latencies. Miozzo and Caramazza took these (null) results as absence of evidence for an input account of the distractor frequency effect. Subsequent studies adopted this conclusion and settled on determining the locus of the distractor frequency effect in the (output) word production stream. Clearly, null findings in a small number of experiments are not sufficient to definitely reject the *input* account and more evidence is needed. The manipulations used in Miozzo and Caramazza's studies (e.g. case alternation) are too specific to be found in studies that did not manipulate them purposely. Word length provides us with a new and alternative way of testing an independent prediction of the input account.

Word length has been shown to influence response times to written words in many studies (Brysbaert et al., 2016). Most models of reading assume that reading can be achieved with one of two pathways, i.e., letter by letter or by mapping the whole word's orthography directly to its meaning. A word-length effect in reading tasks is often taken to suggest that the letters in the words were processed, to some extent, in a sequential fashion (for review, see Barton et al., 2014). Some studies further reported an interaction between word frequency and length. This interaction was taken to suggest that frequent words are more often processed holistically than less frequent words (again, see Barton et al, 2014). Other authors have argued that the longer processing times observed for longer words could be a consequence of the reduced spatial resolution for these words. In longer words, a subset of letters has a higher chance of being less well perceived (Aghababian & Nazir, 2000; Schiepers, 1980). These explanations are not necessarily mutually exclusive and word length effects could have different sources (Frederiksen & Kroll, 1976). Importantly, and unlike for word frequency, the effect of word length has not been theorized to modulate the degree of activation of lexical representations. With this variable, we can test whether a variable that impacts the processing duration of printed distractor words also impacts picture naming latencies in the picture-word interference task. We further examine whether the interaction observed in visual word processing tasks between word length and word frequency is also observed on picture naming latencies in the picture-word interference task. This is as predicted if all the variables that impact distractor processing duration in turn impact target word naming. We note that an effect of distractor length on naming latencies would provide support for the hypothesis that distractor processing duration impacts naming latencies (in line with the input account of the distractor frequency effect) but would not directly clarify the processes underlying the distractor frequency effect itself.

Next, we examine the *interaction between the distractor frequency effect and semantic interference* (i.e., whether the target and distractor and target words are of the same semantic category or unrelated; more details below). This interaction has been deemed crucial to test another account of the distractor frequency, where the effect originates during lexical access for the target word (see Starreveld et al, 2013, and the General discussion for more details). Distractors of the same semantic category as the target word create more interference than unrelated distractors (e.g., Lupker, 1979; Schriefers et al., 1990). This semantic interference effect was and is still often thought to originate during semantic-lexical processing. Miozzo and Caramazza (2003) reasoned that if the distractor frequency effect interacts with this effect, is influenced by the same variables, or has the same time course, this would suggest that the two effects have the same locus. We note right away that an interaction between the two variables would not necessarily be incompatible with accounts in which the two effects have different sources. To describe just one possible scenario, a distractor of low frequency could be processed too late to compete with the target word. If this was the case, only trials with frequent distractors would show a semantic interference effect but this would not mean that the distractor frequency effect arises during lexical access. Miozzo and Caramazza (2003) tested the interaction between distractor frequency and semantic interference in two experiments and did not find supporting evidence. In a later study, Starreveld et al. (2013) observed a significant interaction between the two variables, which they took to support a lexical locus of the distractor frequency effect. With a meta-analysis, we can examine the reliability of this interaction across datasets.

Finally, we examine whether the *frequency of the distractor and that of the target word interact* to modulate naming latencies. The effect of word frequency in simple picture naming tasks is well documented (e.g., Oldfield & Wingfield, 1965; see also Alario et al., 2004; Barry et al., 1997; Jescheniak & Levelt, 1994; Mousikou & Rastle, 2015). Miozzo and Caramazza (2003) tested the interaction between target word and distractor word frequency to determine whether the distractor frequency effect could be accounted for by a *temporal* account. This account is not an explanation of the mechanisms underlying the distractor frequency effect but rather an assessment of the conditions under which the effect arises. However, rejecting this account was deemed crucial by the authors to further constrain their hypotheses about the functional origin of the effect (see their paper and graphs on p. 239 on how a temporal account would be compatible with the competitive account of the distractor frequency effect). In the temporal account, the influence of the distractor depends on the temporal alignment, or synchronization, between target word and distractor word processing. This account implicitly assumes that distractor processing impacts a specific stage of the preparation of the vocal response for the picture. The distractor can only impact the preparation of the target word if processed at a specific time, relative to the processing of the target word (see also Geng et al., 2014). Miozzo and Caramazza reasoned that under this account, the interaction between the properties of the distractor and that of the target word should influence picture naming latencies. In their study, the estimate of the interaction was 3 ms and was not significant. They further argued that under the temporal account, manipulations that should have eased or complicated the recognition of the distractor word in their experiments (e.g., case alternation) should have influenced picture naming latencies. In the light of these results, Miozzo and Caramazza (2003) dismissed the *temporal* account. Recently, Geng et al. (2014) reached a different conclusion. They observed that when participants were made to respond more quickly in the picture-word interference task (e.g., by increasing the number of repetitions and decreasing the number of experimental targets), the distractor frequency effect was no longer present. By contrast when participants' response times were slowed (e.g., by increasing the number of color targets) in the Stroop task, a distractor frequency effect was found. Geng et al. (2014) took these findings to support the temporal account. In their study, Dhooge and Hartuiker (2011) reported an effect size of 18 ms for the interaction between distractor frequency and target word frequency, significant in the by-participant analysis. In the present study, we seek additional evidence in favor of the temporal account by examining again the interaction between distractor and target word frequency.

*Methodological motivations*

The present study will provide information about the properties of the materials that influence naming times in the picture-word interference task. From a methodological perspective, such knowledge is necessary to determine the variables to be controlled in future experiments (a challenge with a long history in psycholinguistics, see Cutler, 1981). This may also prove useful to understand discrepancies across picture-word interference studies, given that different studies tend to use different distractor lists. Contradictory findings are indeed frequent in this literature (e.g., on the role of response set, see Caramazza & Costa, 2000; 2001 or Roelofs, 2001 and on the influence of semantic distance, see e.g., Hutson & Damian, 2014; Mahon et al., 2007; Vieth et al., 2014, to cite a few). In addition, the present study will provide information regarding possible interactions between the properties of the distractor and the properties of the target word. In many picture-word interference studies the manipulation of interest involves the relationship between target and distractor word rather than the properties of the distractor alone. In most of these studies, the same distractor list is used across conditions, but distractors are mapped onto different pictures. This ensures that the effect of the manipulation of interest is not driven by differences in distractor lists across conditions. An interaction between target and distractor word frequency would suggest that the mapping between the properties of the target words and that of the distractors needs to be controlled as well.

Finally, the present study will provide precise estimates of several experimental effects. Such estimates are necessary to conduct a priori power analyses, to assess the probability that an experimental effect is in the wrong direction, or is overestimated (see Gelman & Carlin, 2014). Meta-analytic estimates can be used for this purpose. Precise estimates further allow more accurate computational implementations of interference effects.

In summary, we report a series of Bayesian meta-analyses to quantify the distractor frequency effect and to assess the reliability of various effects and interactions that have shown seemingly discrepant effects in previous research. A Bayesian approach was chosen for two reasons. First, whereas the frequentist approach provides point estimates and an estimate of the uncertainty of the estimator, the Bayesian approach provides a direct estimate of the uncertainty of the parameter of interest; this has the important advantage that we can focus directly on the uncertainty of the estimate. Second, when hypothesis testing becomes necessary (as was the case here), Bayes Factors is superior to frequentist null hypothesis significance testing (NHST) because one can quantify the evidence in favour of the alternative or the null hypothesis given the data; by contrast, in NHST, we can only report the evidence against the null hypothesis.

## Methods

### *Dataset*

We used a subset of the datasets recently collected for a meta-analysis targeting the semantic interference effect (longer naming times with distractors of the same semantic category than for unrelated distractors, Bürki et al., 2020). Bürki et al. (2020) collected a series of datasets corresponding to the following criteria. The task was a classical picture-word interference task. Participants were presented with pictures of objects that they had to name. The pictures were accompanied by distractor nouns. Only the trials where target and distractor were of the same semantic category (e.g., horse-cat), and the corresponding trials where the same targets were associated with unrelated distractors were considered. The distractor was clearly visible and not followed by a mask. Participants were instructed to name the picture as soon as possible after its appearance, and to say the target noun out loud. Participants were all adult speakers without language disorders and the experiments were conducted in the participants' first language. Finally, the dependent variable was the response time for each trial and the raw datasets were available.

Only a subset of these studies was included in the present analyses. The criteria for including a study were: (i) having enough information on the distractors used for each trial, (ii) that the Stimulus Onset Asynchrony (SOA, i.e., time interval between the onset of picture presentation and the onset of distractor presentation) was 0 ms (simultaneous presentation of picture and distractor word), and (iii) that distractors were presented in the written modality (note that all studies with spoken distractors in our database used negative or positive SOAs, so this criterion is subsumed by criterion (ii)). We further restricted the list to studies conducted in an Indo-European language. The final dataset comprised data from 24 different experiments (two of them unpublished). These experiments were further split into 26 separate studies (trials with and without familiarization were treated as separate studies in two experiments). Details about the studies included are provided in Appendix 1. As in Bürki et al. (2020), we followed the procedure described in the original papers to remove incorrect responses, outliers, items, or participants.

Corpora of film subtitles were used to identify the lexical frequency of each distractor and target word[1]. Note that only two datasets included in the meta-analysis come from studies in which the frequency of the distractor was explicitly manipulated, and only five datasets come from studies in which the frequency of the target word was manipulated. Film subtitles were chosen because several studies (Brysbaert, Keuleers, et al., 2011; Cuetos et al., 2011; New et al., 2007) have concluded that frequencies computed from large corpora of subtitles are the best predictors of lexical decision and word naming response times. We used the Subtlex corpora of subtitle frequencies for German (190,500 words; Brysbaert, Buchmeier, et al., 2011), Dutch (437,503 words, Keuleers et al., 2010), Italian (517,564 words), Spanish (94,338 words, Cuetos et al., 2011), US English (74,286 words, Brysbaert & New, 2009), and UK English (160,022 words; Heuven et al., 2014). For French, we used lexeme subtitle frequency from the database *Lexique* (142,694 words, New et al., 2004).

We further counted the number of letters of each distractor and computed or extracted its Orthographic Levensthein Distances (OLD), using the definition in Yarkoni et al. (2008). The OLD between two strings of letters corresponds to the minimum number of insertions, deletions or substitutions required to turn one string into the other. Following standard practice, we computed the OLD20 measure. This consists in first computing the OLD from each distractor word to every other word in a large database, and to then compute the mean OLD of the distractor to its 20 closest orthographic neighbors. Orthographic neighborhood has been shown to influence response times to written words in many studies (Brysbaert et al., 2016). In Ferrand et al. (2018), for instance, frequency, length, and orthographic neighborhood explained over 40% of the variance of response times in a visual lexical decision task. For all languages but Italian, we computed the OLD using the R package "vwr" (Keuleers, 2013), making use of the word databases for several languages (including German, English, Dutch, French and Spanish) provided with the package. For Italian, we took the OLD20 measure provided in the Phonitalia database (Goslin et al., 2014). In the present study, OLD20 is entered as a covariate in the model estimating the effect of lexical frequency. This method provides an estimate of the lexical frequency effect when other variables known to influence visual word processing times are accounted for.

[1] Lexical frequency is often correlated with another variable known to impact picture naming latencies, namely the age at which a word is acquired. Age of acquisition effects (shorter latencies for words acquired earlier) tend to be larger than word frequency effects. Whereas some studies reported no effect of frequency when age of acquisition was controlled for (e.g., Bonin et al., 2002) others reported independent effects of both variables (e.g., Cuetos et al., 1999; Meschyan & Hernandez, 2002). In the present study we used lexical frequency rather than age of acquisition for a practical reason: frequency values were available for all the stimuli in our datasets. Importantly, however, the exact origin of the lexical frequency effect, and whether it is partly driven by differences in age of acquisition is not relevant for the present analysis. Both variables provide a measure of processing speed for the target word.

Descriptive statistics for each variable and dataset are displayed in Appendix 2. A meta-analysis of the effect of the OLD20 measure on picture naming times can be found in Appendix 3.

Word frequency and word length are often correlated (e.g., Piantadosi, 2014). More frequent words tend to be shorter (Zipf, 1935). It is therefore important to assess the independent contribution of each variable. Because we are interested in the effects of word frequency and word length, we would want to report the effects of each of these variables when controlling for the effects of the other one. This can be done by including the two variables in the same statistical model. Estimates for the different variables then can be understood as the contribution of these variables when the variance explained by the other variables has been taken into account. Including all variables in one statistical model only works if the correlations between these variables are not so high that they generate harmful multicollinearity. Figure 1 shows the distribution of the pairwise correlations and Variance

Inflation Factors in the 26 studies. The Variance Inflation Factor (VIF) is computed for each predictor as follows. A linear model is run with the predictor as dependent variable and all other predictors as independent variables. The coefficient of determination ($R^2$) of that model is computed. The VIF is equal to 1/1- $R^2$. If the value is higher than one, it means that this predictor is collinear with the other variables in the model. The criterion to decide whether the Variance Inflation Factor is too high (i.e., will create harmful multicollinearity) varies between authors. Values above 10 are considered a real problem, but several authors consider that values above 4 or 5 can also be problematic (see Hair et al., 2010). As can be seen in Figure 1, the correlations between frequency and word length are moderate and the Variance Inflation Factors are low. The two variables can therefore be entered in the same statistical models.

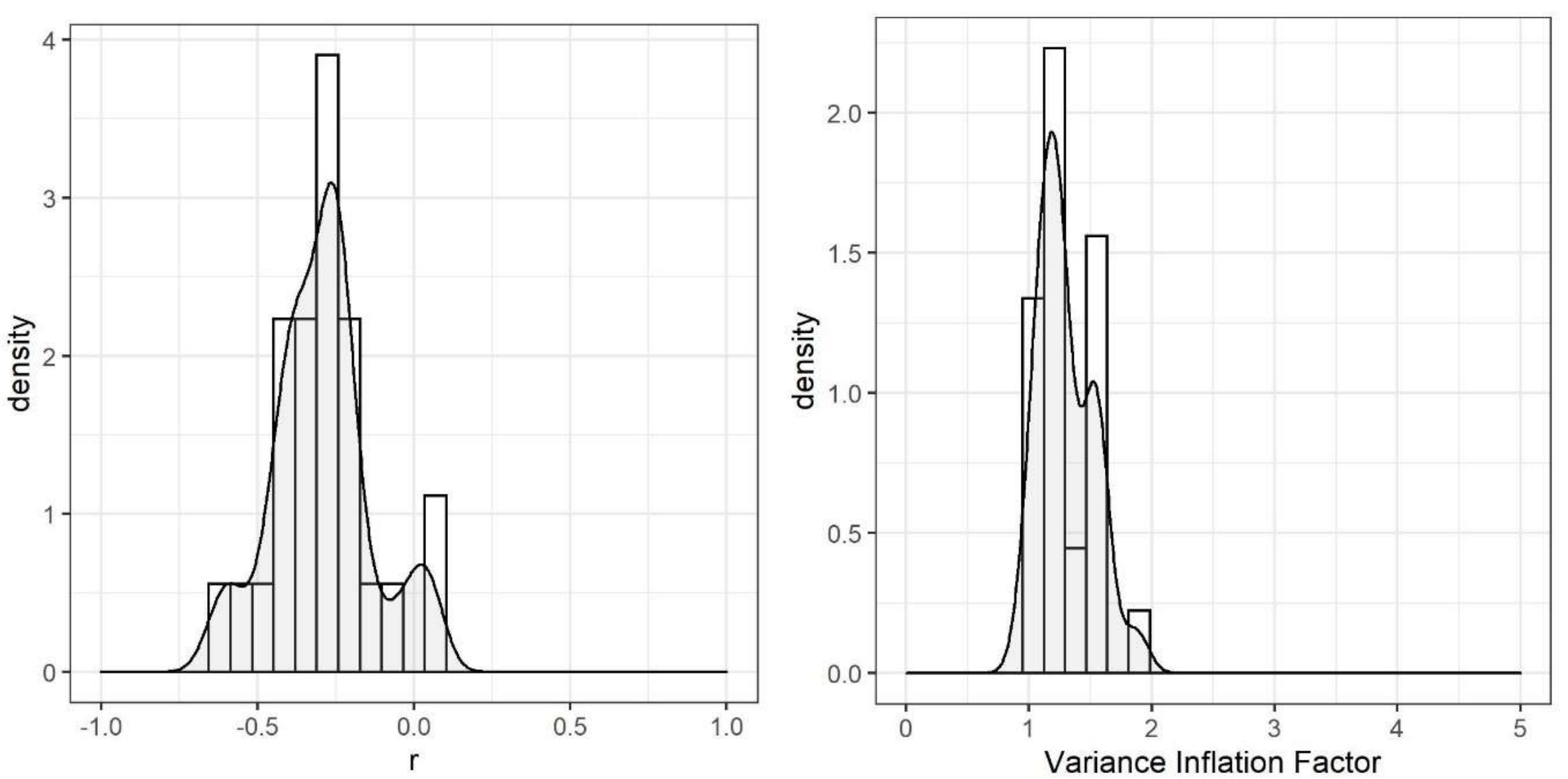

*Figure 1. Distribution of pairwise correlations (Pearson) between distractor frequency and distractor length across datasets (left) and distribution of Variance Inflation Factors (VIFs) across datasets (right)*

### III.3. Meta-analyses

*Extraction of estimates*

The meta-analysis of an effect interest $\theta$ (e.g., increase in response times with each additional letter) takes as input the estimates and standard errors for this effect in a set of individual studies. The output of a meta-analysis is an estimate of this effect and of its uncertainty (usually, a 95% credible interval, which is an interval in which we are 95% certain that the true value of the parameter lies, e.g., Gelman & Carlin, 2014). We conducted several meta-analyses. A summary of these analyses can be found in Table 1. For each analysis, we extracted estimates and standard errors for each dataset, using linear mixed-effects models, as implemented in the library lme4 in R (Bates et al., 2015). In all models, the dependent variable was the untransformed naming time. All models had random intercepts by participant and by item (picture), and all the random slopes allowed by the design. Further specifications are reported in the Results section.

*Table 1. Summary of meta-analyses performed in the present study*

| | **Effect of interest** |
|---|---|
| Meta-analysis 1 | Interaction between distractor frequency and semantic relatedness |
| Meta-analysis 2 | Distractor frequency |
| Meta-analysis 3 | Distractor length |
| Meta-analysis 4 | Interaction between distractor frequency and distractor length |
| Meta-analysis 5 | a. Target word frequency (sanity check)<br>b. Interaction between target frequency and distractor frequency (linear)<br>c. Interaction between target frequency and distractor frequency (quadratic) |

In a first meta-analysis, we examined the interaction between distractor frequency and semantic category (semantically related vs. unrelated distractors). We also examined the interaction between semantic category and distractor length. We started with these analyses to determine whether subsequent analyses should include the interaction with semantic category. In the second meta-analysis, we were interested in the effect of lexical frequency, treated as a continuous measure, when other variables known to also influence processing times for visual words are held constant. In the third and fourth analyses, we examined the effect of word length, once the effect of frequency was accounted for, as well as the interaction between lexical frequency and distractor length. In the last analysis, we focused on the interaction between target word frequency and distractor word frequency. All fixed-effects included in the analyses were centered around the mean.

*Statistical modelling*

Two kinds of meta-analyses can be considered: a fixed-effect (e.g., Chen & Peace, 2013) and a random-effects meta-analysis (e.g., Sutton & Abrams, 2001). Fixed-effects analyses assume the same underlying true effect $\theta$ for all studies while random-effects meta-analyses assume a different true effect $\theta_i$ for each study. In the present study, we opted for the latter because it is reasonable to assume that experiments conducted in different labs and/or in different conditions (e.g., different trial structure, presence or absence of familiarization, different criteria for material selection, experiment length, etc.) have different values for the true effect $\theta_i$.

The meta-analyses were performed under the following assumptions. Each study *i* has an underlying true effect $\theta_i$ that stems from a normal distribution with mean of $\theta$ and standard deviation $\tau$. The observed effect of the predictor $y_i$ in each individual study *i* is assumed to be generated from a normal distribution with mean $\theta_i$ and standard deviation $\sigma_i$ , the true standard error of the study. The model specifications are displayed in Equations (1).

(1)

$$y_i|\theta_i, \sigma_i \sim N(\theta_i, \sigma_i) \; i = 1, \dots, n,$$

$$\theta_i|\theta, \tau \sim N(\theta, \tau),$$

$$\theta \sim N(0, 100),$$

$$\tau \sim N_+(0, 100)$$

$y_i$ is the observed effect of the predictor (i.e., change in the response latencies to the target picture for each unit change in the predictor) in study *i*; $\sigma_i$ is the standard deviation of the estimate from study *i*, estimated from the standard error of the effect of the predictor for this study; $\theta$ is the true effect of this predictor to be estimated by the model; and $\tau$ is the between-study standard deviation. The "+" sign in $N_+$ signals that the normal distribution is truncated to take only positive values.

For each meta-analysis, we first report the meta-analytic estimate, the 95% credible interval and display the posterior distribution computed with weakly informative priors (e.g., Gelman et al., 2013), namely $N(0,100)$ for the intercept and $N_+(0, 100)$ (truncated normal) for standard deviations. We performed the same analysis with two additional priors for the intercept (sensitivity analysis), a normal distribution centered at zero with a standard deviation of 200, and a uniform distribution bounded between -100 and 100. An analysis is consistent with the effect being larger than zero if the 95% CrI does not contain zero. This criterion does not allow us to decide which of the alternative (i.e., the effect is > 0) or null hypothesis (the effect = 0) is more likely given the data. Bayes Factors can provide this information. For each meta-analysis we therefore also computed Bayes Factors. Bayes Factors were set to quantify the evidence for the alternative hypothesis (i.e., there is an effect) over the null hypothesis of no effect. Values larger than one favor the alternative model whereas values below 1 favor the null model. Bayes Factors that are close to one do not provide any concluding evidence. We use Lee and Wagenmakers (2014)'s slightly modified version of Jeffreys' (1961) scale to quantify the evidence in favor of the alternative hypothesis. Bayes Factors between 1 and 3 provide anecdotal evidence, Bayes Factors between 3 and 10 provide moderate evidence, Bayes Factors between 10 and 30 provide strong evidence, and Bayes Factors between 30 and 100 and above 100 provide very strong and extreme evidence, respectively.

The priors on the parameter of interest (here the intercept) represent a priori assumptions about the size of the effect of interest. Bayes Factors can be very sensitive to the priors (e.g., Schad et al., 2022). Weakly informative priors are often not appropriate as they tend to favor the null model (Lee & Wagenmakers, 2014). In the present study, we computed Bayes Factors with four informative priors. These Bayes Factor thus provide information on the evidence in favor of the alternative (over the null) hypothesis given different assumptions about the size of the effect of interest. The dependent variable in the meta-analyses is on the millisecond scale and the properties of the distractors (and target words) are measured on a continuous scale. Thus, the estimate of the meta-analytic model (and its associated prior) represents the increase in milliseconds for each one-unit increase in the variable of interest. The scales of the variables of interest differ and the priors reflect these differences. Distractor word frequency and target word frequency are measured in the natural logarithm of number of occurrences per million. Word length is measured in number of letters, and the computation of the OLD20 measure is reported in the Methods section. For each variable, we selected four different priors, all of which follow a normal distribution and are agnostic with respect to the direction of the effect (the prior distribution is centered at zero). The different priors differ in their standard deviations, and therefore assume a different range of possible effect sizes. To determine appropriate ranges of values for the three variables, we sought information about the effect of these variables in a lexical decision task (participants are presented with sequences of letters and have to decide, by button press, if the sequence forms a real English word or not). We used the data of the English Lexicon Project (ELP, https://elexicon.wustl.edu/, Balota et al., 2007). The ELP dataset contains response times of 820 participants on a total of 40481 words. We restricted the dataset to correct responses to existing words and removed extreme data points (below 150ms and above 3000ms). For each word in the dataset, we counted the number of letters, we extracted the number of occurrences per million from the Subtlex database (Brysbaert & New, 2009) and took the OLD20 measure from the English Lexicon Project database. We took the natural logarithm of the frequency measures. We then centered each of the variables around the mean. We ran a first linear mixed-effects model to estimate the effect of lexical frequency (with OLD20 and number of letters as covariates), a second model to estimate the effect of word length (with lexical frequency as covariate), and a third model to estimate the effect of OLD20 (with lexical frequency as covariate). These estimates are reported in Appendix 4. We can reasonably assume that effects of these same variables on picture naming times in a picture-word interference task are smaller than in tasks where participants respond to the written words directly.

The estimates from the lexical decision task will therefore be used as upper bounds. For each variable, the first informative prior was set to have a standard deviation equivalent to the magnitude of the effect of this variable in the lexical decision dataset. For instance, given an effect size of 45ms for lexical frequency in the lexical decision task, the first prior was a normal distribution with a mean of 0 and a standard deviation of 45. This amounts to assuming that the increase in naming latencies for each one-unit increase in the log of number of occurrences per million for the distractor has a 95% probability of lying between -90ms and 90ms (twice the effect size in the lexical decision task), with small values (i.e., around zero) being more likely than larger values. We then computed Bayes factors for three additional informative priors each time cutting the value of the standard deviation by half. For instance, for lexical frequency we used the additional informative priors $N(0,23)$, $N(0,12)$, $N(0,6)$. For the interactions between semantic category and distractor frequency or distractor length, we used the same priors as for the main effects of the latter variables. For interactions between two continuous variables, we used the priors of the variable with the smallest priors. The priors used for the different variables are listed in Table 2.

*Table 2.* Priors used in the Bayes Factor analyses

| | Prior 1 | Prior 2 | Prior 3 | Prior 4 |
|---|---|---|---|---|
| Distractor word frequency | $N(0,45)$ | $N(0,23)$ | $N(0,12)$ | $N(0,6)$ |
| Word length | $N(0,20)$ | $N(0,10)$ | $N(0,5)$ | $N(0,2.5)$ |
| OLD20 | $N(0,60)$ | $N(0,30)$ | $N(0,15)$ | $N(0,7.5)$ |
| Target word frequency | $N(0,45)$ | $N(0,23)$ | $N(0,12)$ | $N(0,6)$ |
| Semantic interference * Distractor frequency | $N(0,45)$ | $N(0,23)$ | $N(0,12)$ | $N(0,6)$ |
| Semantic interference * Distractor length | $N(0,20)$ | $N(0,10)$ | $N(0,5)$ | $N(0,2.5)$ |
| Distractor frequency * Target frequency | $N(0,45)$ | $N(0,23)$ | $N(0,12)$ | $N(0,6)$ |
| Distractor frequency * Distractor length | $N(0,20)$ | $N(0,10)$ | $N(0,5)$ | $N(0,2.5)$ |

The analyses were performed in R (R Core Team, 2021) using the package “brms” (Bürkner, 2018). Bayes Factors were computed using bridge sampling (Bennett, 1976; Gronau et al., 2017; Schad et al., 2022) with eight chains, 20000 iterations (2000 of these used as warm-up). The datasets and script to reproduce these analyses can be found on OSF (https://osf.io/sjn5b/).

**III. 4. Results**

***Interactions between distractor frequency and semantic relatedness.*** To obtain estimates of the interaction between semantic category and distractor frequency, we fit a mixed-effects model with semantic category, distractor frequency, and their interaction as fixed-effects. We then used the estimates and standard errors of the interaction terms to fit a Bayesian meta-analysis. The meta-analysis estimate is of 0.3ms with a 95% CrI ranging from -3.4 to 3.8ms. The posterior distribution of the between-study standard deviation has a mean of 3.25ms (CrI: 0.2, 8.6). The results of sensitivity analyses for the present and following meta-analyses are presented in Appendix 5. The posterior distribution of this interaction as well as Bayes Factors are plotted in Figure 2. The 95% Credible interval contains zero and Bayes factors provide moderate to strong evidence in favor of the null model.

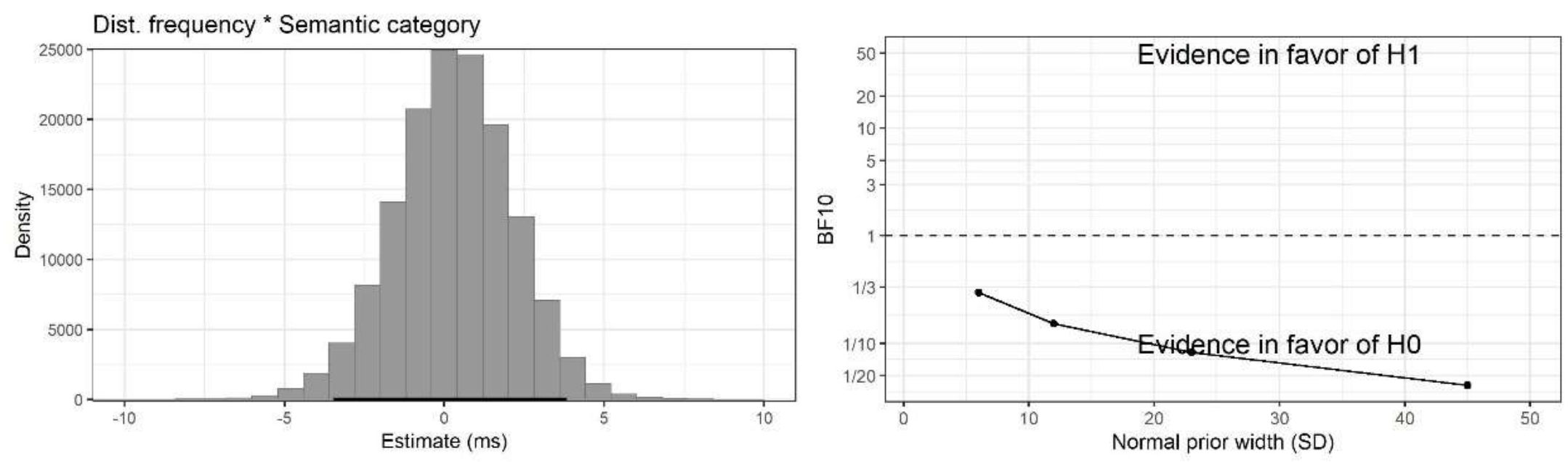

*Figure 2. left:* Posterior distribution for the estimate of the interaction between semantic category and distractor frequency; *right*: Bayes Factors

This analysis does not provide support for the hypothesis that semantic category interacts with distractor frequency. Rather, there is moderate evidence against this interaction. Following the same procedure, we also conducted a meta-analysis of the interaction between semantic category and distractor length. The results of this analysis are similar, they are presented in Appendix 5. In all subsequent analyses, we included trials with related and unrelated distractors and did not include interactions with semantic relatedness.

***Distractor frequency.*** To obtain estimates of the distractor frequency effect in individual studies, we conducted, for each of them, a mixed-effects model with the logarithm of lexical frequency, the number of letters, and the orthographic Levensthein distance (OLD 20) as fixed effects. We then performed a meta-analysis with the individual estimates of distractor frequency. The meta-analysis reveals that the overall effect of distractor frequency is of -4.4ms with a 95% CrI ranging from -7.2 to -1.9ms. According to the sign of the estimate, as the frequency of the distractor increases, picture naming latencies decrease. Given that the frequency measure is on the natural log scale, adding one unit amounts to multiplying the number of occurrences per million by 2.72. This means that two distractors whose frequencies in number of occurrences per million differ by a factor of 2.72 (e.g., 10 vs. 27.2 occurrences per million, or 100 vs. 272 occurrences per million) lead to a naming time difference of 4.4 ms).

The posterior distribution of the between-study standard deviation has a mean of 2.2ms (CrI: 0.1, 5.8). The posterior distribution of the distractor frequency estimate and Bayes factors are plotted in Figure 3a. Bayes factors with informed priors all provide strong evidence in favor of the alternative model. Figure 4 displays the posterior distributions of the estimates of the distractor frequency effect for each study.

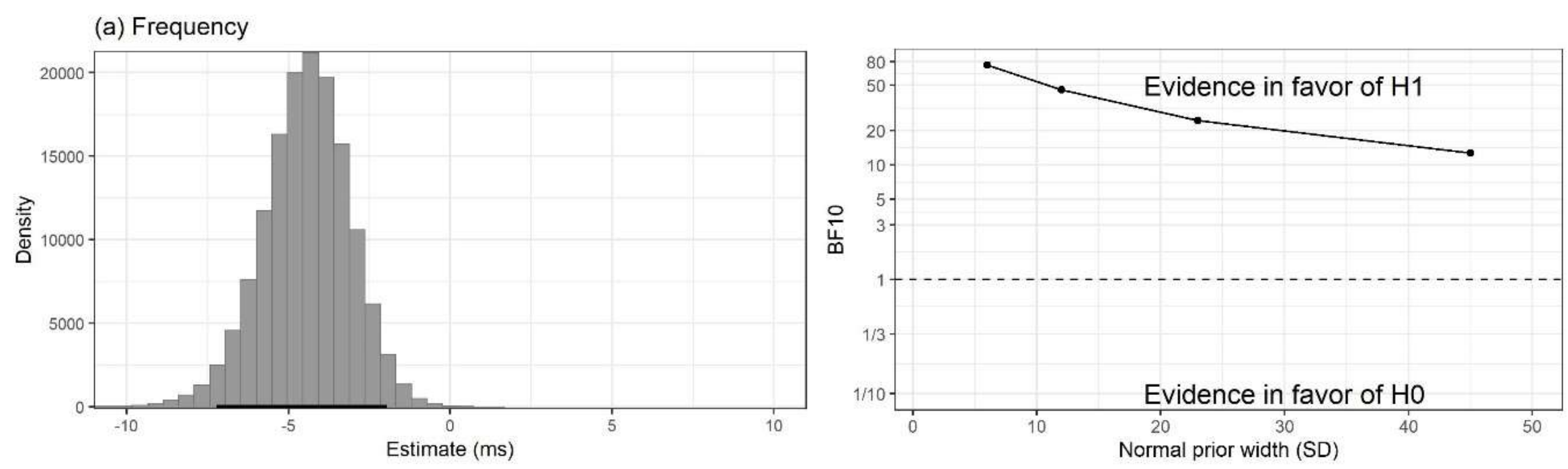

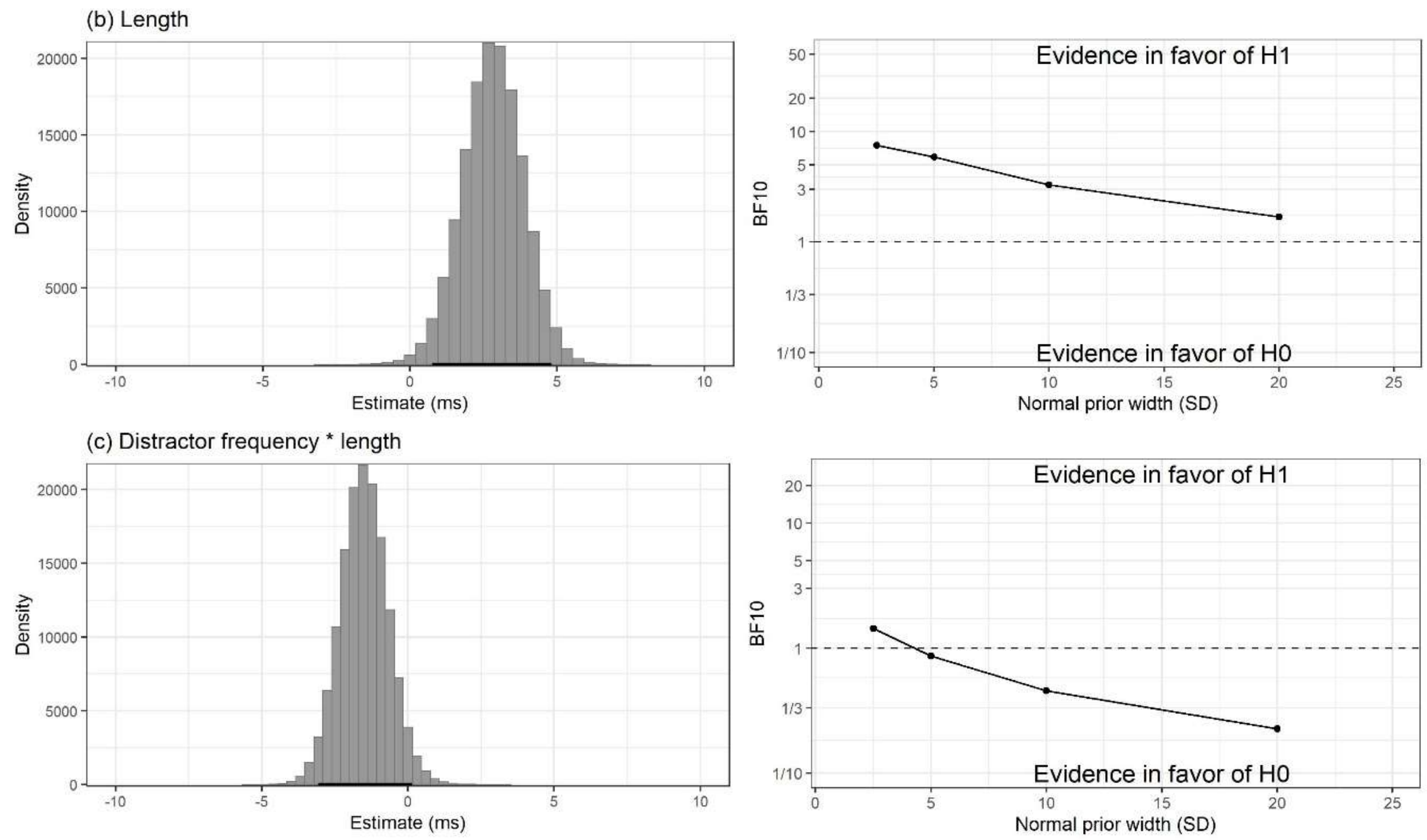

*Figure 3.* Left: Posterior distributions of the estimate of (a) distractor frequency (b) distractor length and (c) the interaction between distractor frequency and distractor length with 95% Credible Interval (black horizontal line); Right: corresponding Bayes Factors

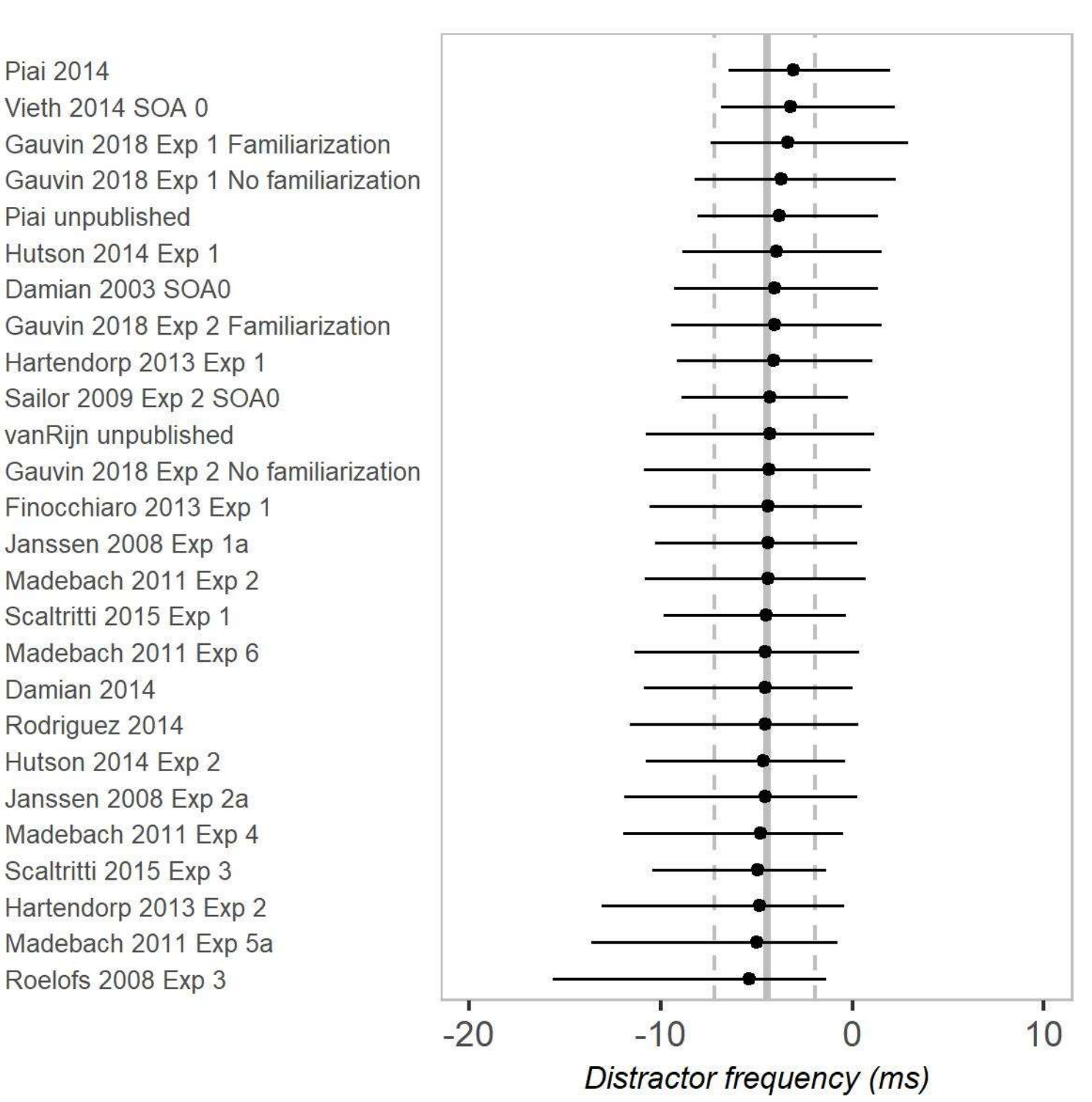

*Figure 4. Summary of the random-effects meta-analysis modelling the effect of distractor frequency on naming times. For each study, the figure displays, in black, the mean and posterior estimate (mean and 95% credible interval). A negative value means that distractors with higher frequency result in shorter naming latencies for the picture. The grey vertical line represents the grand mean (i.e., the meta-analytic effect) and the dashed vertical lines delimit the 95% credible interval of that estimate*

***Distractor length.*** To obtain estimates of the distractor length effect in individual studies, we conducted a mixed-effects model with number of letters and lexical frequency as fixed-effects, for each study. We then used the estimates for the variable number of letters as input for the meta-analysis. This analysis reveals that the overall effect of distractor length is about 2.8ms (CrI: 0.8 , 4.8). Trials with longer distractors are named with longer naming latencies. The posterior distribution of the between-study standard deviation has a mean of 1.5ms (CrI: 0.06, 4.3). The posterior distribution of the distractor length estimate as well as Bayes Factors are plotted in Figure 3b. Bayes factors with the three more constraining priors (standard deviations of 10, 5, or 2.5) provide moderate support for the alternative model whereas the Bayes Factor with the less constraining prior is inconclusive. Figure 5 displays the posterior distributions of the estimates of the distractor length effect for each study.

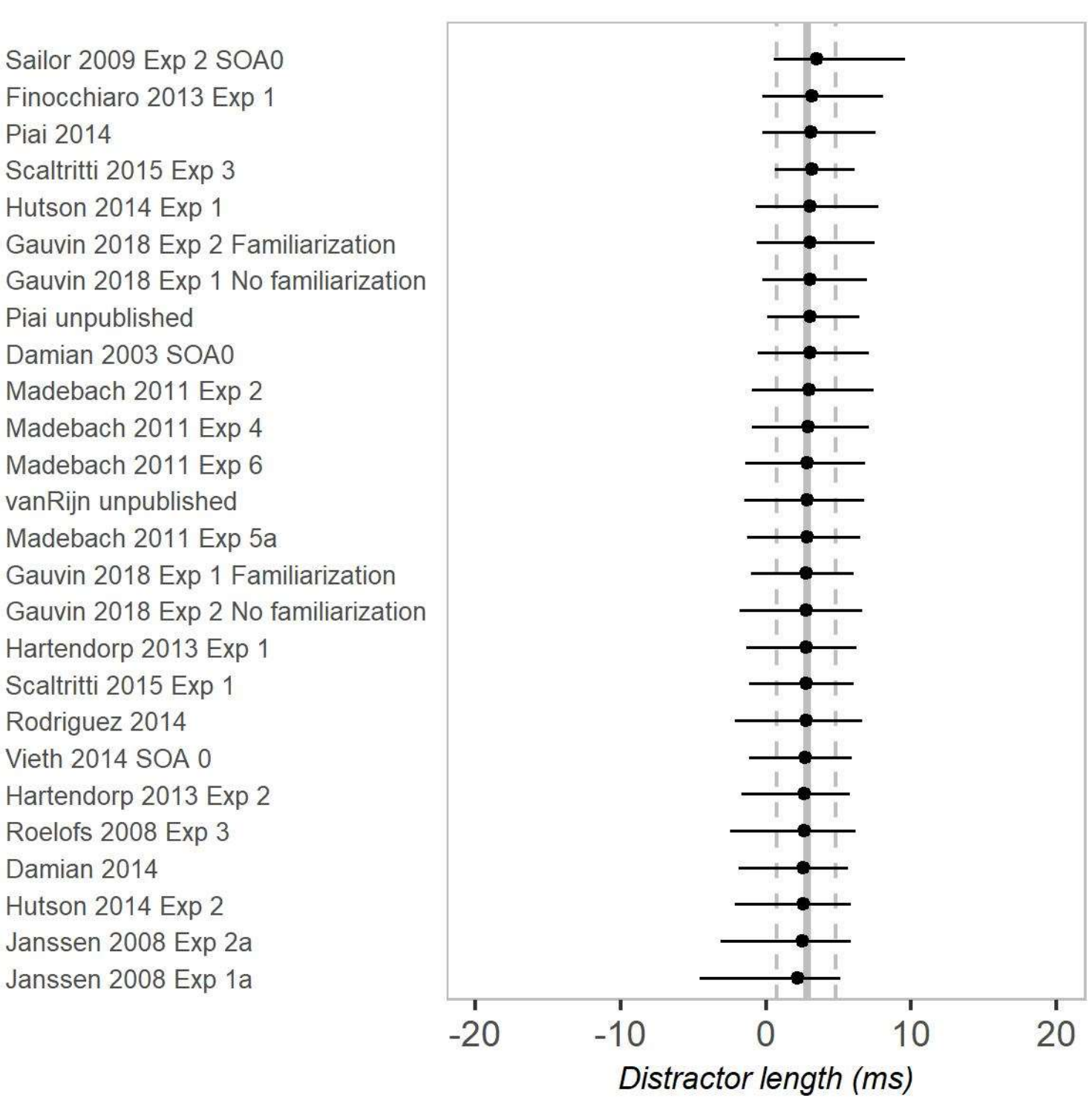

*Figure 5. Summary of the random-effects meta-analysis modelling the effect of distractor length on naming times. For each study, the figure displays, in black, the mean and posterior estimate (mean and 95% credible interval). A positive value means that distractors with more letters result in longer picture naming latencies. The grey vertical line represents the grand mean (i.e., the meta-analytic effect) and the dashed vertical lines delimit the 95% credible interval of that estimate.*

***Interaction between lexical frequency and distractor length.*** To obtain estimates of the interaction between the frequency of the distractor and the length of that distractor, we conducted, for each study, a linear mixed-effects model with main effects of distractor frequency, word length and their interaction. The meta-analytic estimate for the interaction term is about -1.5ms (CrI: -3.1, 0.1). The sign of the interaction suggests that the effect of distractor length is smaller for high than for low frequency words. The posterior distribution of the between-study standard deviation has a mean of 1.1ms (CrI: 0.04, 3.2). The posterior distribution and Bayes factors are displayed in Figure 3c. A visual representation of the posterior distributions of the estimates of this interaction for each study, can be found in Appendix 6. With a standard deviation for the prior of 20ms, the Bayes factor provides moderate support for the null hypothesis. With standard deviations of 10 or smaller, the evidence does not favor either hypothesis.

***Interaction between distractor frequency and target frequency.*** As a sanity check, we conducted a meta-analysis of the main effect of target word frequency. A facilitative effect of target word frequency is a staple feature of picture naming performance that has been reported in many previous studies (e.g., Oldfield & Wingfield, 1965; see also Alario et al., 2004; Barry et al., 1997; Ellis & Morrison, 1998; Jescheniak & Levelt, 1994; Mousikou & Rastle, 2015). It was important to show that we could replicate this effect with our datasets before testing the interaction between target word frequency and distractor word frequency. To extract estimates of this effect in individual studies, we conducted a mixed-effects model with the logarithm of target word frequency as fixed-effect. These estimates were then used as input for the meta-analysis. The result of this meta-analysis supports the hypothesis that more frequent target words are named with shorter naming times. The overall effect is about -7.9ms (95% CrI: -11.5, -4.8), Bayes Factors provide very strong to extreme evidence in favor of the alternative hypothesis. The posterior distribution of the target frequency estimate and Bayes Factors are plotted in Figure 6a. A visual representation of the posterior distributions of the estimates of the target word frequency effect for each study, can be found in Appendix 7. The posterior distribution of the between-study standard deviation has a mean of 4.1ms (CrI: 0.3, 8.7).

*Interaction term (linear)*. To obtain estimates of the interaction between the frequency of the distractor and the frequency of the target word in the individual studies, we conducted, for each study, a linear mixed-effects model with the main effects of target and distractor word frequencies and an interaction term. We then used the individual estimates and Standard errors of the interaction to perform a meta-analysis. This analysis reveals that the meta-analytic estimate for the interaction term is about 2.2ms (95% CrI: 0.5 , 4.1). The sign of the meta-analytic estimate suggests that the effect of distractor frequency decreases as the frequency of the target word increases. To better understand this interaction, we computed the meta-analytic estimates of the intercept and main effects of target word and distractor frequency in the model with the interaction. These are respectively 764 (SE=31.7), -7.55 (SE=1.46) and -7.38 (SE=1.61) ms. We then entered these values in the regression equation to compute predicted response times at different values of the target and distractor word frequencies, taking, for illustration purposes, the target and distractor word frequency distributions from one study in our database (Gauvin et al. 2018, no familiarization). In this study, the mean target frequency in number of occurrences per million was 6337 (SD = 6638) and the mean distractor frequency was 9681 (SD = 11483). We log transformed and centered the frequency values and used them to compute the predicted effect of distractor frequency (i.e., predicted difference in naming times for two values of distractor frequency that differ by one standard deviation) for three values of target word frequency, namely the mean, -1 SD and +1 SD. When target frequency is 1 SD below the mean, the distractor frequency effect is 12 ms. When target frequency is at its mean, the distractor frequency effect is 9ms, and when target frequency is 1 SD above the mean, the distractor frequency effect is 5 ms.

The posterior distribution of the between-study standard deviation has a mean of 1.6ms (CrI: 0.07, 4.2). The posterior distribution of the meta-analytic estimate for the interaction and Bayes Factors are plotted in Figure 6b. Bayes Factors favor the alternative model (with the interaction) when the effect size has a prior with SD = 6ms. For larger SDs, Bayes factors are inconclusive. A visual representation of the posterior distributions of the estimates of this interaction for each study can be found in Appendix 8.

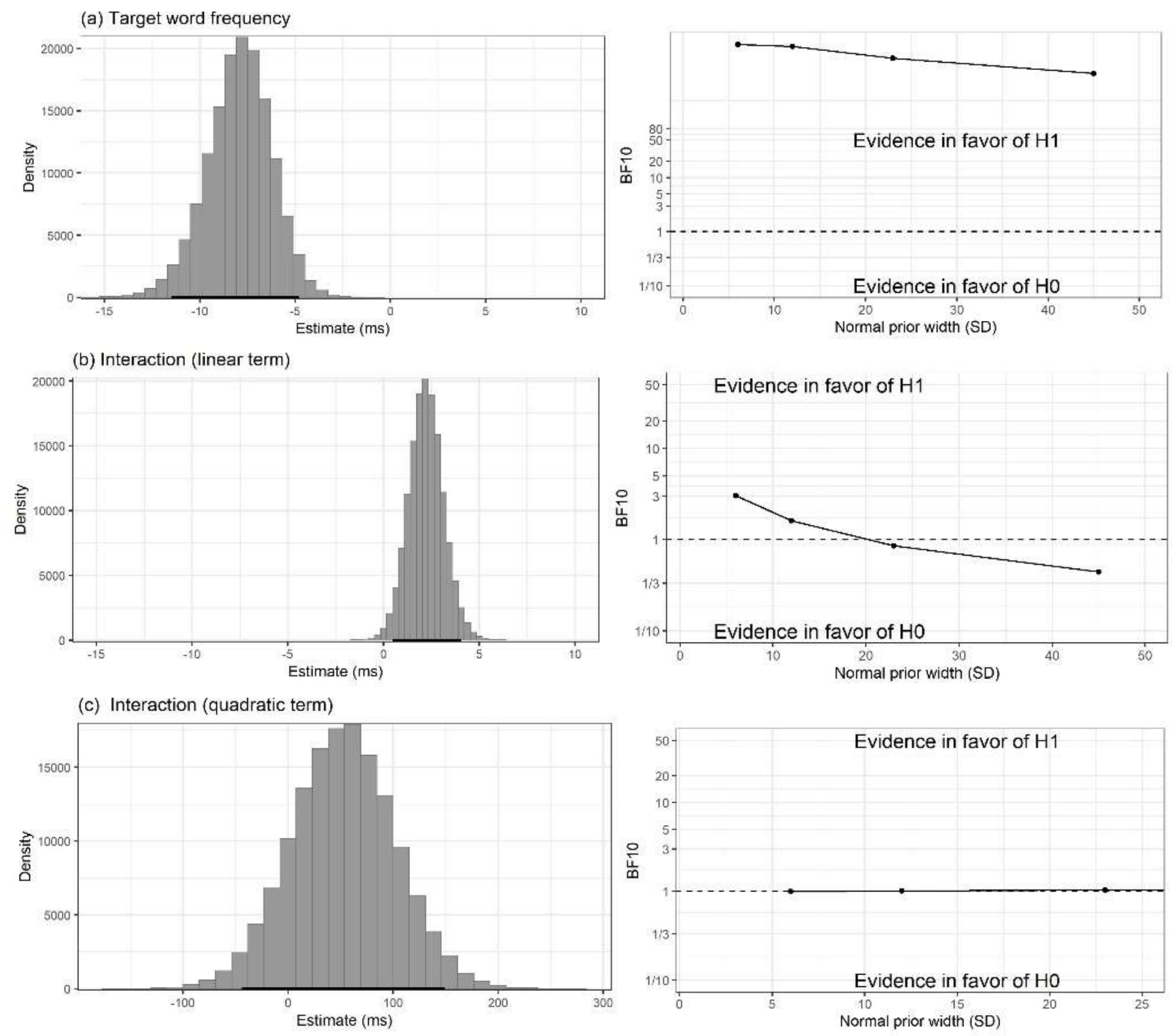

*Figure 6.* Posterior distributions of the estimate and Bayes Factors for (a) target word frequency, (b) the interaction term between target word frequency and distractor word frequency (linear term), (c) the interaction between target word frequency and distractor word frequency (quadratic term).

*Interaction term (quadratic)*. For this analysis, we conducted, for each study, a linear mixed-effects model with target word frequency, linear and quadratic terms for distractor word frequency, and interactions between target word frequency and both the linear and quadratic terms for the variable distractor frequency. We then performed a meta-analysis of the interaction between target word frequency and the quadratic term. The meta-analytic estimate for the interaction term is about 52.5ms. The 95% Credible interval contains zero (-43.8 , 148.9). The posterior distribution of the between-study standard deviation has a mean of 61.7ms (CrI: 2.4 , 174.9). The posterior distribution of the meta-analytic estimate and Bayes Factors are plotted in Figure 6c. Bayes Factors do not provide any conclusive evidence. A visual representation of the posterior distributions of the estimates of this interaction for each study, can be found in Appendix 9.

**General Discussion**

The present study investigated the influence of the properties of distractor and target words on picture naming times in picture-word interference tasks. We conducted a series of meta-analyses with estimates extracted from 26 datasets.

Our first aim was to quantify the size of the distractor frequency effect in experiments where this variable is not manipulated. Previous studies in which this effect was studied compared groups of high frequency and low frequency distractors. Our analyses confirmed the reliability of the distractor frequency effect (i.e., longer naming times for trials with low frequency distractors) and revealed that this effect is also detected when the variable *frequency* is not dichotomized, nor purposely manipulated (all but two datasets in the meta-analysis are from studies in which the frequency of the distractor was not manipulated). Our meta-analysis thus provides information on the size of this effect in typical picture-word interference experiments. A one-unit increase in frequency (logarithm of number of occurrences per million scale) decreases response times to the pictures by about 4.4ms [-7, -2]. If we use this estimate to compute the predicted response times for frequency values that correspond to the mean frequencies of the low and high frequency categories in Miozzo and Caramazza (2003), we obtain an effect of about 20ms, a value that is very similar to the value reported in their experiments. In the present study, the estimates were computed in models including other distractor properties as predictors, namely word length and orthographic neighborhood. As a result, our analysis is informative about the size of the effect of distractor frequency when these other variables are controlled for.

Our second aim was to examine the impact of a novel variable, distractor length, on picture naming times, as a test of the hypothesis that distractor processing times impact naming latencies in the picture-word interference paradigm. The analysis suggests that pictures accompanied by shorter distractors are named more quickly than pictures accompanied by longer distractors. The estimate of the meta-analysis suggests that naming times increase by 2.8ms with each additional letter in the distractor. Hence, everything else being equal, participants will produce a given target word with the distractor word *butterfly* 17ms slower than the same target word with the distractor word *cat*. The results regarding the interaction between lexical frequency and distractor length are less conclusive. The estimate of this interaction has the same sign as the interaction observed in visual word processing tasks, with a larger effect of distractor length for low frequency words than for high frequency words. The 95% Credible interval overlaps with zero and the Bayes Factor analysis does not provide evidence that the alternative hypothesis is more likely than the null hypothesis given the data.

The third aim of the present study was to examine the interaction between distractor frequency and semantic category. This interaction was deemed important by several authors (Miozzo & Caramazza, 2003; Starreveld et al., 2013) to determine the locus of the distractor frequency effect. Our data do not provide support in favor of this interaction. Rather, Bayes Factors favor the null hypothesis.

The last aim of our study was to test the reliability of the interaction between distractor frequency and target word frequency. Previous studies reported inconsistent findings. Our analyses replicate the target word frequency effect in picture naming (i.e., faster response times to pictures with more frequent nouns) and provide some evidence that word frequency and target word frequency interact. The distractor frequency effect decreases when target word frequency increases. Bayes Factor analyses provide moderate evidence that the effect is larger than zero when assuming that the effect lies between -12 and 12ms. Bayes Factors with less constraining priors are inconclusive. Note that because we did not control for the age of acquisition of the target words, a variable often correlated with lexical frequency, we do not know whether the target frequency effect is a pure frequency effect or is partly driven by an age of acquisition effect. The exact origin of this effect is however not relevant for the present analysis. The frequency variable provides a measure of the speed with which the target word can be prepared for production.

In the remainder of this discussion, we examine the theoretical and methodological implications of these findings in the context of the available accounts of the distractor frequency effect and more generally, for picture-word interference studies.

*Could the distractor frequency effect result from input processes?*

After Miozzo and Caramazza (2003) found a “reverse” distractor frequency effect, they performed several experiments to determine whether this effect was due to input processes alone (i.e., visual word recognition of the written word) or whether it reflected the processing of the target word (output processes). Based on a series of null effects, they rejected the input account and concluded that the distractor frequency effect reflects (and informs on) word production processes. Most subsequent studies adopted this assumption and set out to determine the locus of the distractor frequency effect within the word production system, without re-considering the input account as a viable account. The impact of distractor length provides support for the hypothesis that the duration of distractor processing influences naming times. To our knowledge, this is the first demonstration that a variable that influences the duration of distractor processing (and not potentially also its activation level) impacts picture naming latencies in the picture-word interference paradigm. One possible interpretation of this finding is that distractor processing duration modulates picture naming latencies in this paradigm. This observation can lead to the generalized hypothesis that any variable that impacts visual word recognition processes, including distractor frequency, will influence naming times. The results of the present study thus suggest that the input account may have been rejected too hastily.

Miozzo and Caramazza also rejected the hypothesis that the distractor frequency effect arises because of differences in the temporal alignment of target and distractor word processes for frequent and infrequent distractors. In this “temporal” account, the distractor can only impact the preparation of the target word if processed at a specific time, relative to the processing of the target word. Miozzo and Caramazza reasoned that under this account, the frequency of the target and distractor words should interact but there was no such interaction in their data. The interaction that we observe between target word frequency and distractor frequency is in line with the temporal account. It is further in line with results reported by Geng et al. (2014), who observed that the distractor frequency effect disappeared when participants were made to respond more quickly.

As discussed in the Introduction, the input and temporal accounts are not orthogonal. Any variable that influences the duration of distractor or target word processing will in turn affect the temporal alignment of their processing. The interaction between distractor and target word frequencies is compatible with an account in which interference effects result from a decrease in available cognitive resources, the impact of which depends on how much word production processes overlap with the processing of the distractor. Accordingly, if the production processes have already been completed when the processing of the distractor begins, the distractor will have less of an impact on picture naming latencies.

Another observation of the present study is the absence of evidence for an interaction between distractor frequency and semantic relatedness. This interaction has been used as a test case to determine the locus of the distractor frequency effect, and more specifically, to decide between a lexical and non-lexical locus of this effect. In competitive models of lexical access, the semantic interference effect originates in lexical access. Starreveld et al. (2013), and before them Miozzo and Caramazza (2003), reasoned that if the distractor frequency effect also arises during lexical access, the two variables should interact. The present meta-analysis does not provide support for this interaction. Starreveld et al. (2013) took the absence of interaction in their data to signal a different locus for the semantic interference and distractor frequency effects. If we would follow the same logic here, several loci remain possible for the distractor frequency effect, including the hypothesis that this effect is an input effect.

The findings of the present study are thus compatible with the hypothesis that the distractor frequency effect is an input effect. Notably, other findings in the literature can easily be explained by an input account. Dhooge and Hartsuiker (2011) reported a distractor frequency effect in a delayed version of the picture-word interference naming task. The participants were asked to name the picture upon seeing the distractor word, which appeared 1000ms after the picture. Dhooge and Hartsuiker (2011) observed a distractor frequency in this task but no effect of the frequency of the target word. These results are expected under the input account. Recall that this account assumes that the processing of the distractor takes some of the resources that would otherwise be used to process the target word. Upon seeing the distractor, and for the duration of distractor processing, less resources will be available to process the target word. Consequently, processing time for this word will increase and this increase will reflect the time needed to process the distractor. In another study, Dhooge & Hartsuiker (2010) compared trials where the distractor was masked (i.e., the distractor was presented for a very brief period of time, and preceded and followed by a sequence of hash marks or randomly selected sequence of consonants) to trials where the distractor was fully visible. The distractor frequency effect was only found when the distractor was fully visible. Masking the distractor has been argued to result in a more superficial treatment of this word (see for instance Roelofs et al., 2011). If the distractor is not fully processed, it is less likely to take up processing resources and delay the processing of the target word. Another finding consistent with an input account of the distractor frequency is the interaction reported by Miozzo and Caramazza (2003) between distractor frequency and phonological relatedness. Distractors that overlap phonologically with the target word generate less interference than unrelated distractors (e.g., Bi et al., 2009; Damian & Martin, 1999; Posnansky & Rayner, 1977; Rayner & Posnansky, 1978; Starreveld & La Heij, 1996;  de Zubicaray et al. 2002). The phonological facilitation effect has been shown to depend on the timing of presentation or processing between target and distractor word (Damian & Martin, 1999, Schriefers et al.; 1990; Bürki, 2017). If, as our data suggest, the frequency of the distractor modulates the temporal alignment between distractor and target word processing, an interaction between distractor frequency and phonological overlap can be expected[2].

Miozzo and Caramazza (2003) rejected the input account of the distractor frequency effect. Many other researchers adhered to this conclusion. Given the findings of the present study, we argue that this conclusion needs to be reconsidered. To be clear, our conclusion that the estimated effect is compatible with an input account does not imply that it is not, or that it is less compatible with other accounts of the effect. Moreover, the observation that distractor processing duration impacts naming latencies does not allow us to conclude that the distractor frequency is only due to differences in processing duration between frequent and infrequent words.

*Alternative explanations of distractor frequency effects?*

Most of the debate on the functional origin of the distractor frequency effect in the last two decades involved two opposing views, the post-lexical and the lexical view. According to the post-lexical view, also termed the *response exclusion* hypothesis (e.g., Finkbeiner & Caramazza, 2006; Mahon et al., 2007), speakers unwillingly prepare the distractor word for production and store the motor response in a pre-articulatory buffer. This account further assumes that only one word can be in the response buffer at a given point in time, therefore, to produce the target word, the speaker must first empty this buffer (see also Dhooge & Hartsuiker, 2010; 2011). According to this account, the interference effect depends on the time at which the distractor enters the buffer. If it enters the buffer earlier, it can be suppressed earlier. It follows that the distractor should only create interference if it can enter the response buffer before target word processing has reached a certain processing stage. The quicker the target word is processed relative to the distractor word, the lower the probability that the distractor reaches the response buffer before the target word, yielding a weaker interference effect

[2] The same reasoning applies for the semantic interference effect, as this effect also depends on the timing of presentation of target and distractor words (e.g., Schriefers et al., 1990). Importantly, however, whether an interaction between the semantic (or phonological) effect and distractor frequency is observed will depend on the participants processing times for target and distractor words in the experiment (see Bürki & Madec, in press).

(but see Starreveld et al, 2013, for a different view). Hence, in this account, variables that influence the relative timing of distractor and target word processing (and therefore, any variable that influences either the speed and/or duration of target word or distractor word processing) are expected to modulate naming times.

According to the alternative lexical view, the distractor frequency effect arises during lexical access and is therefore informative of this process. For instance, Starreveld et al. (2013) build on the assumption –made in several models of reading – that lexical representations for more frequent words have lower selection thresholds than more frequent words (see also Besner & Risko, 2016). Starreveld et al. (2013) further assume that following recognition, the distractor representation returns to its resting activation level, with a decay rate proportional to its activation level. It follows that in the picture-word interference task, high frequency distractor words reach lower activation levels and, consequently, act as weaker competitors yielding faster response times. In this account, distractor frequency impacts naming times because of differences in *degrees of activation* between frequent and less frequent distractors. Our finding that word length impacts naming times shows that a variable that does not influence the degree of activation also modulates naming times. Starreveld et al. (2013) should thus make the additional hypothesis that the distractor influences naming latencies via both processing duration and activation levels. It must be stressed however that an input account of the distractor frequency effect is not incompatible with the idea that lexical access is a competitive process. For instance, the Weaver++ model assumes that lexical access is a competitive process. It further assumes that in the picture-word interference task, the speaker accesses the lexical representation associated with the distractor word, which then has to be blocked. Logically, the distractor only has to be blocked when the target lexical representation has not yet been selected by the time the distractor lexical representation becomes activated. The speed with which the blocking mechanism can take place has been argued to depend on the speed with which the distractor is processed (Roelofs, 2005; Roelofs et al., 2011).

To summarize, the present study provides novel insights regarding the conditions in which interference from an unrelated distractor word occurs. Our primary conclusion is that the meta-analytic estimates we report challenge the rejection of the input account, without compromising alternative accounts. The upshot is that the interpretation of the distractor frequency effects is still uncertain. We argue that the available evidence does not suffice to associate the distractor frequency effect with genuinely output word production processes, or to determine the mechanisms underlying the distractor frequency effect. Subsequent studies will need to consider the input account when designing experiments to pinpoint the functional origin of the distractor frequency effect and its general consequences for word production processes.

*Methodological implications*

Before concluding, we highlight the methodological implications of the present findings. While the majority of picture-word interference studies report comparisons involving the same list of distractors counterbalanced across conditions, not all of them do (e.g., Bürki et al., 2019; Finkbeiner & Caramazza, 2006; Foucart et al., 2010; Mahon et al., 2007; Rizio et al., 2017). The finding that naming latencies are influenced by word length and by orthographic neighborhood (see Appendix 3) calls for a strict control of these properties if different distractor lists are to be used. One strategy that is often used to "control for" these variables is to make sure that the distractor lists for different conditions do not differ statistically. This strategy is not optimal, because differences across lists, even non-significant, can have a significant impact on the dependent variable, when tested in a group of participants (see Sassenhagen & Alday, 2016, for a detailed discussion). A better strategy might be to control for these variables in the statistical model, as we did here.

More generally the finding that the properties of written words known to influence visual processing times also influence naming times in the picture-word interference task could explain part of the discrepancies across studies in the picture-word interference literature (e.g., the conditions in which the semantic interference effect surfaces, see Bürki et al. 2020 for review; or the differences in the Stimulus Onset Asynchronies at which experimental effects are observed) given that different studies usually use different material lists and may involve different languages. The interaction between the properties of the distractor and target words further suggests that, even when identical target and distractor lists are used across conditions (e.g., in studies on the semantic interference effect or on the phonological facilitation effect, e.g., Damian & Martin, 1999; Posnansky & Rayner, 1977; Rayner & Posnansky, 1978), differences across conditions can be expected that are not driven by the experimental manipulations of interest (i.e. target-distractor relationships) but by differences in the temporal alignment between target and distractor word processing.

Finally, the meta-analytic estimates of the present study can be used to guide future experimental studies. Precise estimates of experimental effects are necessary to conduct a priori power analyses, to assess the probability that an experimental effect is in the wrong direction, or is overestimated (see Gelman & Carlin, 2014). Effect sizes extracted from meta-analyses are particularly valuable given that they consider a large sample of the available data. Often, meta-analytic estimates tend to overestimate the true effect. This is because papers that end up getting published will tend to not include work from experiments where the experimental effect did not reach significance (only significant results tend to be published). The estimates of the present study are likely to suffer less from this limitation given that, they focus on experimental effects that were not the primary focus of the published papers (two datasets come from studies in which distractor frequency was manipulated and five from studies in which target word frequency was manipulated; none of the studies manipulated distractor length or orthographic neighbourhood). We provide effect sizes and precision estimates for several factors of primary interest: distractor frequency, distractor length, target word frequency, as well as for the interaction between target word frequency and distractor word frequency, distractor word frequency and semantic relatedness, and distractor frequency and distractor length. These estimates might also prove useful to guide further specifications of existing accounts in future work, for instance to simulate naming times for different target-distractor combinations in the picture-word interference paradigm under different scenarios. They will also serve as a useful summary of existing data to further refine computational models that make quantitative predictions about the magnitude and range of effects in the picture-word interference paradigm (e.g., Weaver++, Roelofs, 2003) or to implement new computational models. To our knowledge, only the distractor frequency effect has been simulated so far, a simulation inspired by effect sizes obtained in experiments where the variable frequency was dichotomized. Our meta-analytic estimates can be used to extend these models. In Weaver++ parameters are set such that trials with high frequency distractors are produced 25msec faster than trials with low frequency distractors (Roelofs, 2005; Roelofs et al., 2011). The meta-analytic estimate of distractor frequency that we provide informs on the change in naming times with each unit change in distractor frequency and can be used to make more nuanced predictions.

## Conclusion

We conducted a series of meta-analyses of the effects of distractor properties on naming times in the picture-word interference paradigm. The results of these analyses suggest that it is too early to reject an input account of the distractor frequency effect. More generally, the available data are compatible with this as well as with other accounts of the effect. In our view, strong conclusions about the mechanism(s) underlying the distractor frequency may have been reached prematurely.

## Acknowledgements

The authors would like to thank all the authors who shared their datasets. This research was funded by the Deutsche Forschungsgemeinschaft (DFG, German Research Foundation) – project number 317633480 – SFB 1287, Project B05 (Bürki) and project Q (Vasishth/Engbert). F.-Xavier Alario was supported by grants ANR-16-CONV-0002 (ILCB) and the Excellence Initiative of Aix-Marseille University (A*MIDEX).

**Declaration of Conflicting Interests**

The Authors declare that there is no conflict of interest

*Appendix 1. Datasets included in meta-analyses*

| **Study ID (graphs)** | **Experiment n° in paper** | **Reference** |
|---|---|---|
| Mädebach 2011 Exp. 2 | Experiment 2 | Mädebach et al. (2011) |
| Mädebach 2011 Exp. 4 | Experiment 4 | |
| Mädebach 2011 Exp. 5a | Experiment 5a | |
| Mädebach 2011 Exp. 6 | Experiment 6 | |
| Janssen 2008 Exp. 1a | Experiment 1a | Janssen et al. (2008) |
| Janssen 2008 Exp. 2a | Experiment 2a | |
| Scaltritti 2015 Exp. 1 | Experiment 1 | Scaltritti et al. (2015) |
| Scaltritti 2015 Exp. 3 | Experiment 3 | |
| Piai unpublished | - | - |
| Damian 2014 | - | Damian & Spalek (2014) |
| Gauvin 2018 Exp. 1 Familiarization | Experiment 1 | Gauvin et al. (2018) |
| Gauvin 2018 Exp. 1 No familiarization | Experiment 1 | |
| Gauvin 2018 Exp. 2 Familiarization | Experiment 2 | |
| Gauvin 2018 Exp. 2 No familiarization | Experiment 2 | |
| Damian 2003 SOA0 | - | Damian & Bowers (2003) |
| Piai 2014 | - | Piai et al. (2014) |
| Vieth 2014 SOA 0 | | Vieth et al. (2014) |
| Hartendorp 2013 Exp. 1 | Experiment 1 | Hartendorp et al. (2013) |
| Hartendorp 2013 Exp. 2 | Experiment 2 | |
| Hutson 2014 Exp. 1 | Experiment 1 | Hutson & Damian, (2014) |
| Hutson 2014 Exp. 2 | Experiment 2 | |
| Roelofs 2008 Exp. 3 | Experiment 3 | Roelofs (2008) |
| vanRijn unpublished | - | - |

| Rodriguez 2014 | - | Rodríguez-Ferreiro et al. (2014) |
| --- | --- | --- |
| Sailor 2009 Exp.2 SOA0 | Experiment 2 | Sailor et al. (2009) |
| Finocchiaro 2013 Exp. 1 | Experiment 1 | Finocchiaro & Navarrete (2013) |

*Appendix 2. Descriptive statistics (Mean, Standard Deviation and Range) for the variables distractor frequency, length, OLD20 in each individual study*

| Study | Frequency | | | | Length | | | | OLD20 | | | |
|---|---|---|---|---|---|---|---|---|---|---|---|---|
| | Mean | SD | Min | Max | Mean | SD | Min | Max | Mean | SD | Min | Max |
| Damian 2003 SOA0 | 25.93 | 22.24 | 2.52 | 68.40 | 4.56 | 1.17 | 3 | 6 | 1.45 | 0.40 | 1 | 2.45 |
| Damian 2014 | 31.81 | 70.30 | 0.55 | 321.82 | 5.24 | 1.37 | 3 | 9 | 1.82 | 0.23 | 1.3 | 2.5 |
| Finocchiaro 2013 Exp. 1 | 5770.96 | 6263.46 | 583.00 | 25238.00 | 5.61 | 1.10 | 4 | 8 | 1.59 | 0.24 | 1 | 1.95 |
| Gauvin 2018 Exp. 1 Familiarization | 9281.16 | 10908.99 | 384.00 | 55649.00 | 4.98 | 1.75 | 3 | 10 | 1.78 | 0.76 | 1 | 3.8 |
| Gauvin 2018 Exp. 1 No familiarization | 9681.16 | 11482.61 | 384.00 | 55649.00 | 4.90 | 1.76 | 3 | 10 | 1.75 | 0.74 | 1 | 3.8 |
| Gauvin 2018 Exp. 2 Familiarization | 4675.00 | 5675.20 | 284.00 | 32275.00 | 5.27 | 1.41 | 3 | 9 | 1.78 | 0.49 | 1 | 3.35 |
| Gauvin 2018 Exp. 2 No familiarization | 4756.63 | 6011.74 | 284.00 | 32275.00 | 5.22 | 1.47 | 3 | 9 | 1.77 | 0.51 | 1 | 3.35 |
| Hartendorp 2013 Exp. 1 | 111.87 | 418.50 | 0.21 | 2465.94 | 5.46 | 1.64 | 2 | 8 | 1.77 | 0.62 | 1 | 2.85 |
| Hartendorp 2013 Exp. 2 | 81.96 | 126.20 | 0.96 | 458.00 | 5.78 | 2.56 | 3 | 11 | 1.90 | 0.80 | 1 | 3.8 |
| Hutson 2014 Exp. 1 | 3271.98 | 3644.02 | 63.00 | 22361.00 | 5.19 | 1.58 | 3 | 10 | 1.79 | 0.60 | 1 | 3.85 |
| Hutson 2014 Exp. 2 | 2908.65 | 3442.10 | 31.00 | 11628.00 | 5.13 | 0.86 | 3 | 7 | 1.84 | 0.39 | 1 | 2.5 |
| Janssen 2008 Exp. 1a | 594.45 | 690.91 | 57.00 | 3716.00 | 5.48 | 1.27 | 3 | 9 | 1.95 | 0.48 | 1.3 | 3.45 |
| Janssen 2008 Exp. 2a | 38.24 | 58.12 | 0.83 | 255.28 | 5.98 | 1.15 | 3 | 8 | 1.94 | 0.35 | 1 | 2.75 |
| Mädebach 2011 Exp. 2 | 14.37 | 32.00 | 0.24 | 162.41 | 5.45 | 1.29 | 4 | 8 | 1.75 | 0.37 | 1.05 | 2.7 |
| Mädebach 2011 Exp. 4 | 10.83 | 14.71 | 0.24 | 70.79 | 5.33 | 1.11 | 3 | 8 | 1.69 | 0.37 | 1 | 2.5 |
| Mädebach 2011 Exp. 5a | 2023.86 | 1864.20 | 128.00 | 8886.00 | 5.48 | 1.33 | 3 | 10 | 1.95 | 0.48 | 1.3 | 3.45 |
| Mädebach 2011 Exp. 6 | 2024.01 | 1876.48 | 128.00 | 8886.00 | 5.48 | 1.32 | 3 | 10 | 1.95 | 0.47 | 1.3 | 3.45 |
| Piai 2014 | 159.33 | 652.86 | 1.99 | 4025.85 | 4.55 | 1.19 | 3 | 7 | 1.49 | 0.48 | 1 | 2.7 |
| Piai unpublished | 179.69 | 691.18 | 1.99 | 4025.85 | 4.81 | 1.59 | 3 | 9 | 1.58 | 0.58 | 1 | 3.1 |
| Rodriguez 2014 | 20.31 | 28.46 | 0.05 | 129.42 | 5.72 | 1.29 | 4 | 8 | 1.89 | 0.46 | 1 | 2.95 |
| Roelofs 2008 Exp. 3 | 40.96 | 102.06 | 0.02 | 458.00 | 6.19 | 1.29 | 4 | 9 | 2.03 | 0.56 | 1.05 | 3.1 |
| Sailor 2009 Exp. 2 SOA 0 | 1571.74 | 3164.75 | 21.00 | 16831.00 | 5.29 | 1.22 | 3 | 8 | 1.91 | 0.70 | 1 | 3.75 |
| Scaltritti 2015 Exp. 1 | 1705.10 | 2425.42 | 32.00 | 13494.00 | 6.81 | 1.75 | 4 | 11 | 2.05 | 0.62 | 1 | 3.8 |
| Scaltritti 2015 Exp. 3 | 1685.59 | 2393.16 | 32.00 | 13494.00 | 6.82 | 1.75 | 4 | 11 | 2.05 | 0.62 | 1 | 3.8 |
| vanRijn unpublished | 41.03 | 101.02 | 0.02 | 458.00 | 6.31 | 1.42 | 4 | 9 | 2.04 | 0.56 | 1.05 | 3.1 |
| Vieth 2014 SOA 0 | 1758.88 | 2294.19 | 25.00 | 11117.00 | 5.89 | 1.49 | 3 | 9 | 2.00 | 0.67 | 1 | 3.5 |
| | | | | | **5.50** | **1.43** | **3.23** | **8.89** | **1.83** | **0.52** | **1.05** | **3.17** |

*Appendix 3. Results of meta-analysis of distractor orthographic Levensthein distance (OLD20)*

| *Prior* | *Estimate* | *CrI (95%)* |
|---|---|---|
| Normal (0,100) | 11.6 | [5.4 , 17.4] |
| Normal (0,200) | 11.6 | [5.4 , 17.4] |
| Uniform (-100,100) | 11.6 | [5.4 , 17.4] |

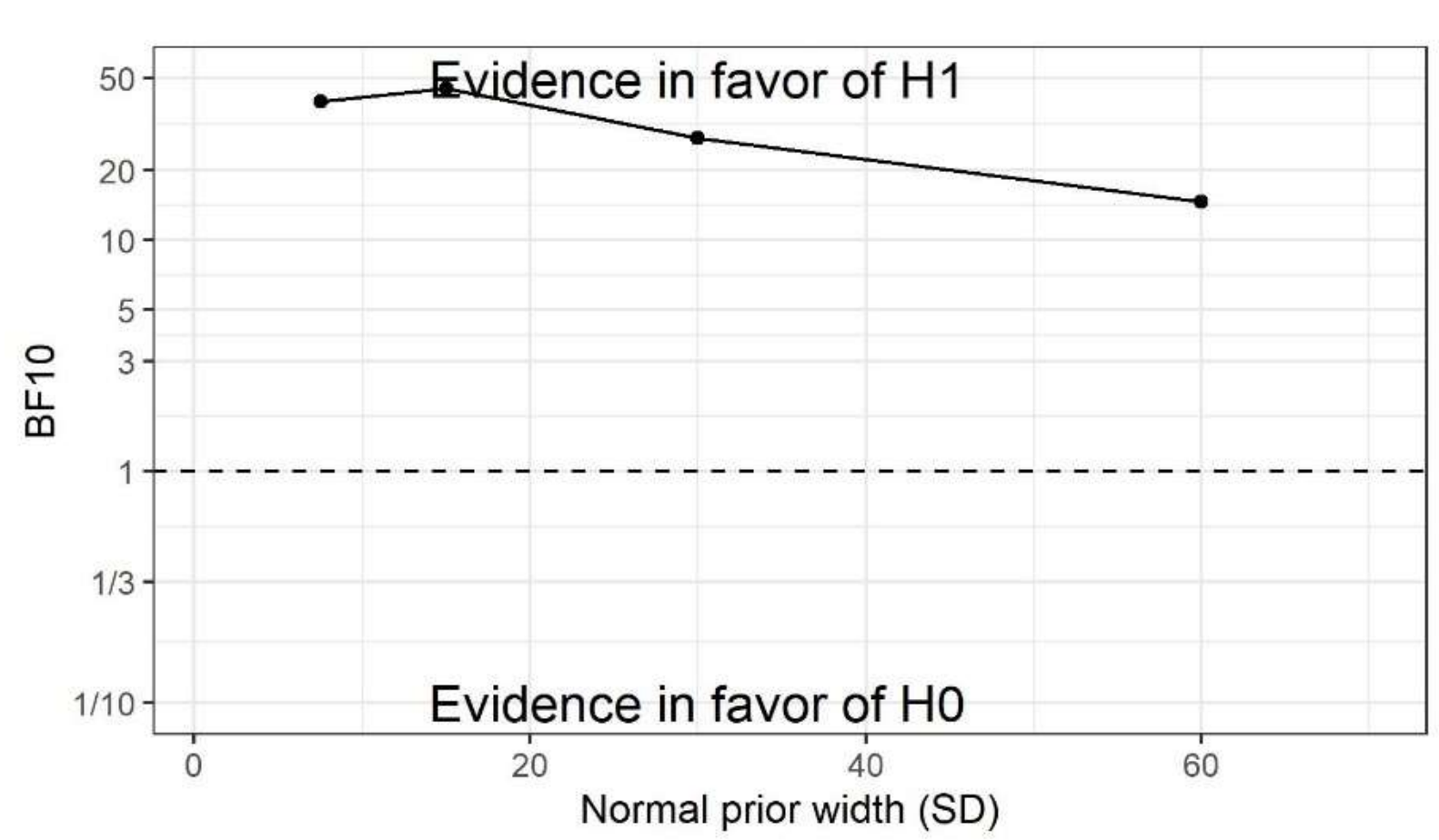

*Appendix 4. Effects of word frequency, number of letters and OLD20 on lexical decision times in the English Lexicon Project*

| *Independent variable* | *Estimate* | *SE* |
|---|---|---|
| Lexical frequency (log number of occurrences per million) | -45 | 1.06 |
| Word length (Number of letters) | 21 | 0.64 |
| OLD20 | 58 | 1.54 |

*Appendix 5. Sensitivity analyses. Meta-analytic estimates and 95% Credible Intervals with different priors for the effects of distractor frequency, distractor length, target word frequency, the interaction between distractor frequency and target word frequency (linear and quadratic terms), and the interaction between distractor frequency and distractor length*

| Effect of distractor frequency | | |
|---|---|---|
| *Prior* | *Estimate* | *CrI (95%)* |
| Normal (0,100) | -4.45 | [-7.20 , -1.96 ] |
| Normal (0,200) | -4.45 | [-7.20 , -1.96] |
| Uniform (-100,100) | -4.01 | [-6.41 , -1.87] |

| Effect of distractor length | | |
|---|---|---|
| *Prior* | *Estimate* | *CrI (95%)* |
| Normal (0,100) | 2.81 | [0.75 , 4.81] |
| Normal (0,200) | 2.81 | [0.75 , 4.82] |
| Uniform (-100,100) | 2.81 | [0.73 , 4.83] |

| Effect of target word frequency | | |
|---|---|---|
| *Prior* | *Estimate* | *CrI (95%)* |
| Normal (0,100) | -7.94 | [-11.54 , -4.78] |
| Normal (0,200) | -7.93 | [-11.50 , -4.80] |
| Uniform (-100,100) | -7.94 | [-11.54 , -4.78] |

| Interaction target word * distractor frequency (linear) | | |
|---|---|---|
| *Prior* | *Estimate* | *CrI (95%)* |
| Normal (0,100) | 2.22 | [0.46 , 4.07] |
| Normal (0,200) | 2.22 | [0.46 , 4.06] |
| Uniform (-100,100) | 2.22 | [0.45 , 4.06] |

| Interaction target word * distractor frequency (quadratic) | | |
|---|---|---|
| *Prior* | *Estimate* | *CrI (95%)* |
| Normal (0,100) | 52.47 | [-43.75 , 148.93] |
| Normal (0,200) | 64.34 | [ -41.58, 170.72] |
| Uniform (-100,100) | 42.18 | [ -46.87, 97.06] |

| Interaction distractor frequency *distractor length | | |
|---|---|---|
| *Prior* | *Estimate* | *CrI (95%)* |
| Normal (0,100) | -1.49 | [-3.07 , 0.14] |
| Normal (0,200) | -1.49 | [-3.07 , 0.15] |
| Uniform (-100,100) | -1.49 | [-3.08 , 0.15] |

| Interaction semantic category * distractor frequency | | |
|---|---|---|
| *Prior* | *Estimate* | *CrI (95%)* |
| Normal (0,100) | 0.28 | [-3.46 , 3.83] |
| Normal (0,200) | 0.29 | [-3.42 , 3.83] |
| Uniform (-100,100) | 0.29 | [-3.46 , 3.89] |

| Interaction semantic category * distractor length | | |
|---|---|---|
| *Prior* | *Estimate* | *CrI (95%)* |
| Normal (0,100) | 1.21 | [-2.9 , 5.43] |
| Normal (0,200) | 1.22 | [-2.89 , 5.48] |
| Uniform (-100,100) | 1.22 | [-2.89 , 5.44] |

*Appendix 6. Forest plot, meta-analysis of the interaction between Distractor frequency and Distractor length*

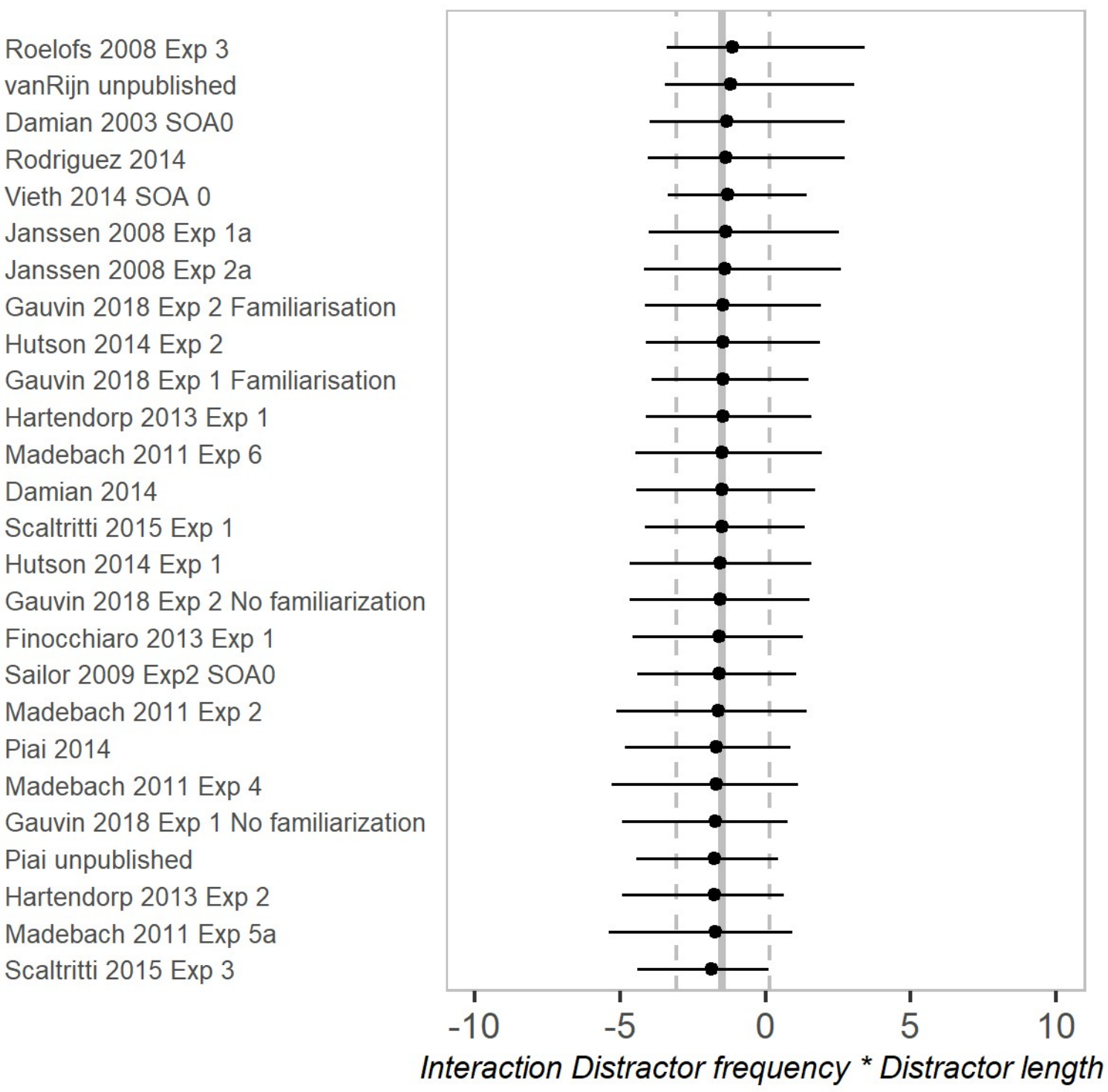

*Summary of the random-effects meta-analysis modelling the interaction between distractor word frequency and distractor length on naming times. For each study, the figure displays, in black, the mean and posterior estimate (mean and 95% credible interval). A negative value means that the influence of word length is smaller for high frequency words. The black vertical line represents the grand mean (i.e., the meta-analytic effect) and the dashed vertical lines delimit the 95% credible interval of that estimate.*

*Appendix 7. Forest plot, meta-analysis of the main effect of target word frequency*

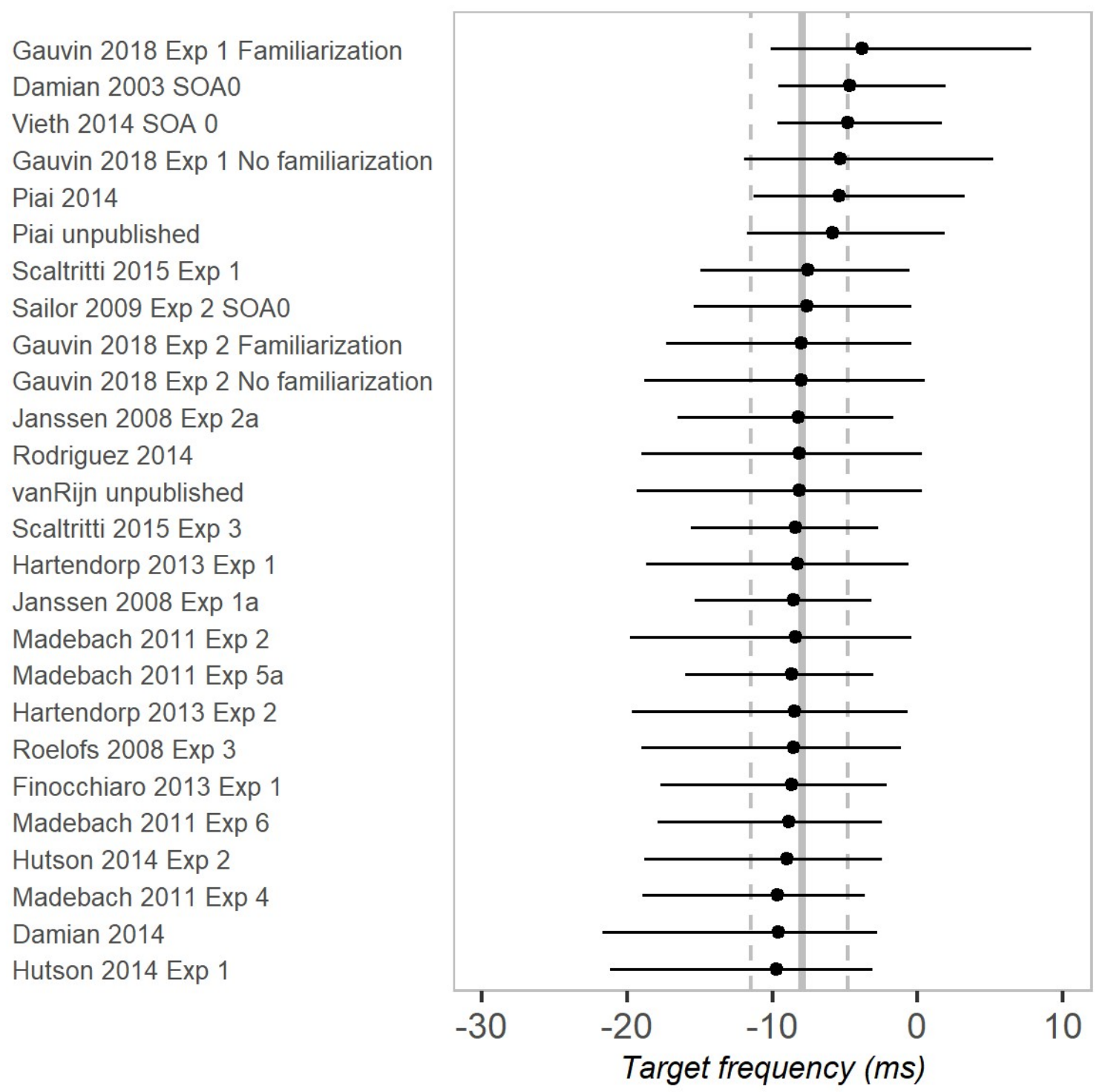

*Summary of the random-effects meta-analysis modelling the effect of target word frequency on naming times. For each study, the figure displays, in black, the mean and posterior estimate (mean and 95% credible interval). A negative value means that distractors with a greater frequency value result in shorter naming latencies for the picture. The grey vertical line represents the grand mean (i.e., the meta-analytic effect) and the dashed vertical lines delimit the 95% credible interval of that estimate.*

*Appendix 8. Forest plot, meta-analysis of the interaction between Distractor frequency (linear term) and Target word frequency*

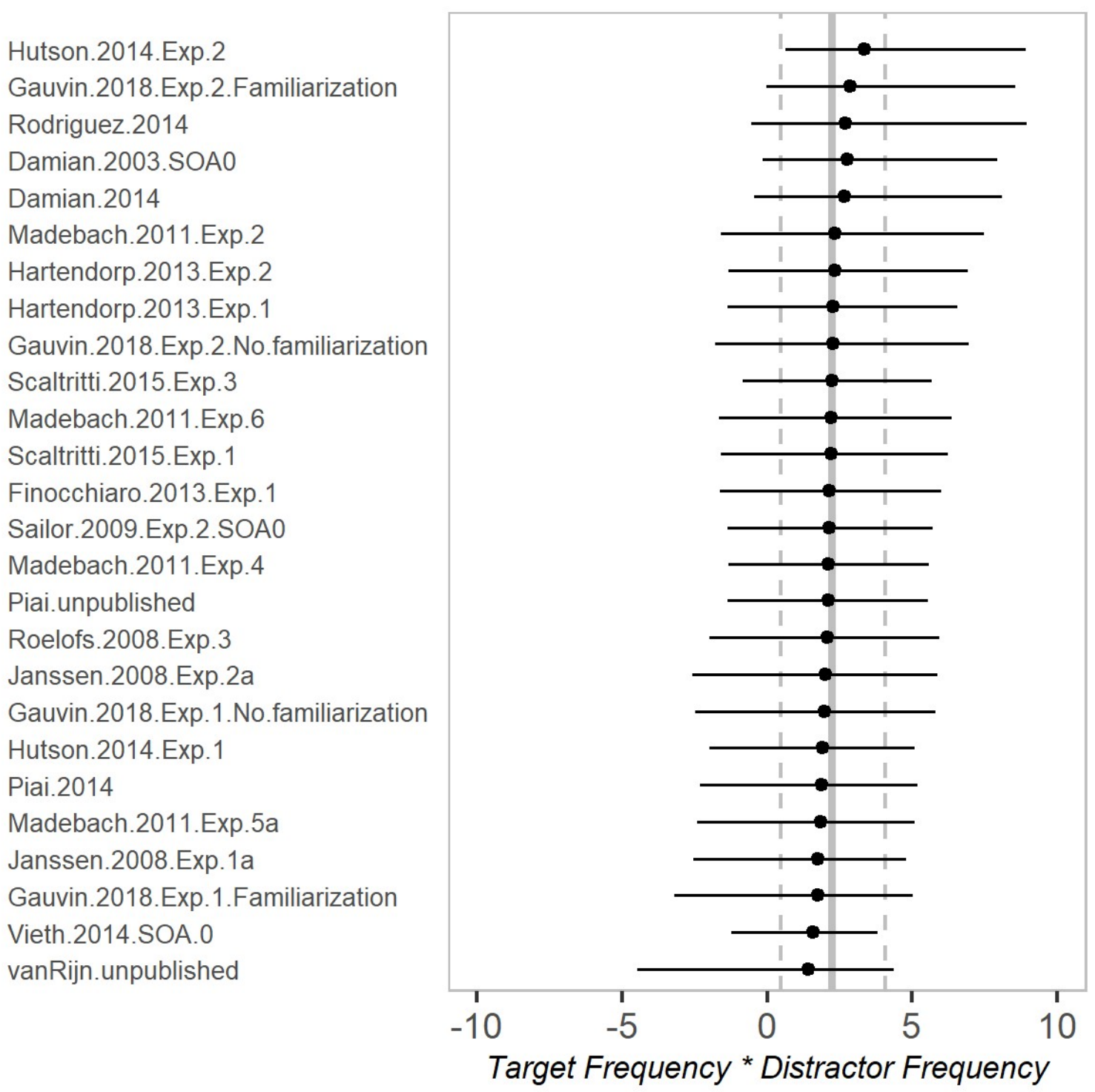

*Summary of the random-effects meta-analysis modelling the interaction between target word frequency and distractor word frequency on naming times. For each study, the figure displays, in black, the mean and posterior estimate (mean and 95% credible interval). A positive value means that target words with a higher frequency value show less of a facilitative effect of distractor frequency. The grey vertical line represents the grand mean (i.e., the meta-analytic effect) and the dashed vertical lines delimit the 95% credible interval of that estimate.*

*Appendix 9. Forest plot, meta-analysis of the interaction between Distractor frequency (quadratic term) and Target word frequency*

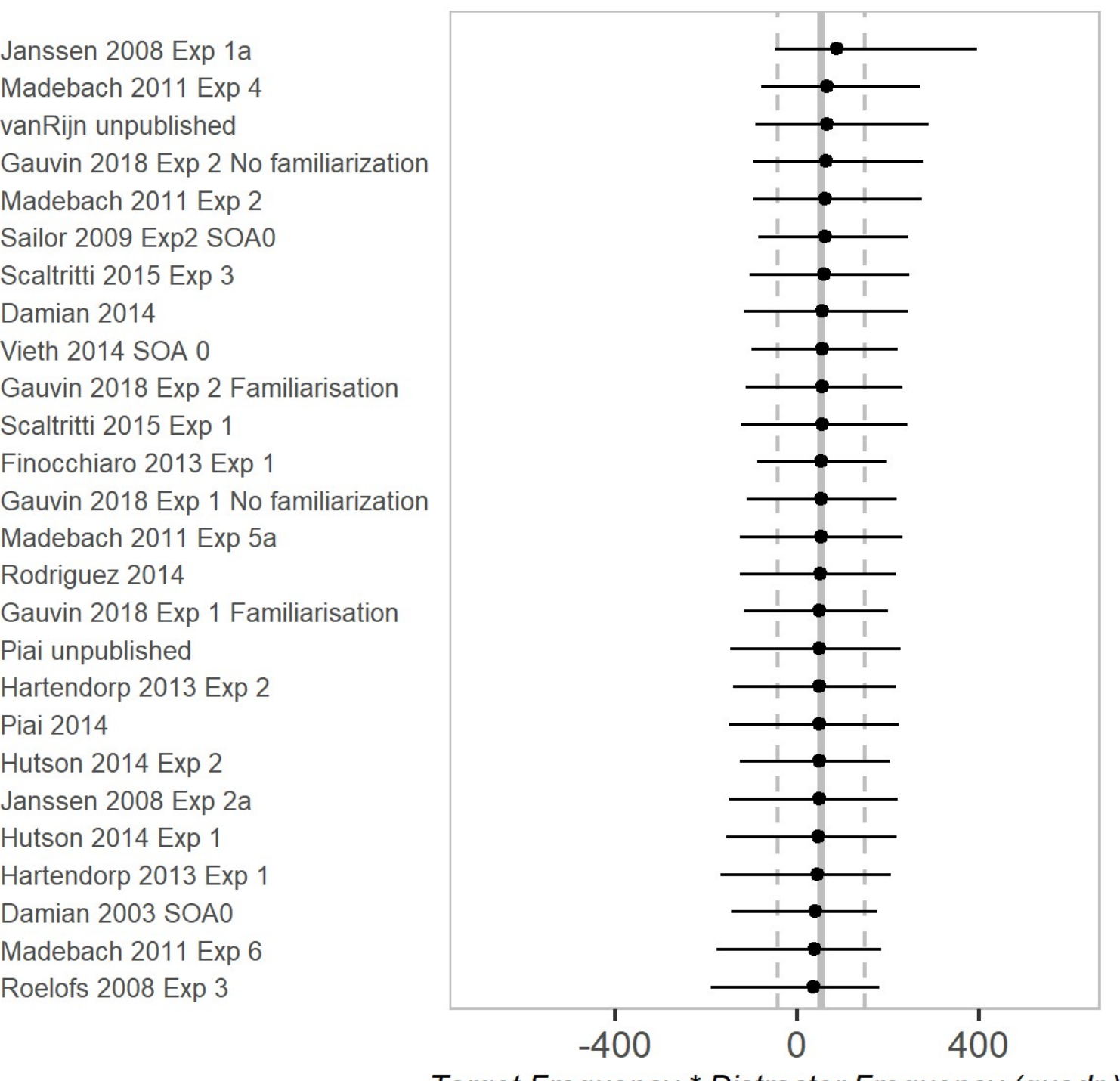

*Summary of the random-effects meta-analysis modelling the interaction between target word frequency and distractor word frequency (quadratic term) on naming times. For each study, the figure displays, in black, the mean and posterior estimate (mean and 95% credible interval). The grey vertical line represents the grand mean (i.e., the meta-analytic effect) and the dashed vertical lines delimit the 95% credible interval of that estimate.*